%% file: HHvbf.tex
\documentclass[pdftex,smallextended,epjc3]{svjour3}
\pdfoutput=1
\RequirePackage[T1]{fontenc}
\usepackage[left=2.5cm,right=2.5cm,top=2.5cm,bottom=2.5cm]{geometry}
\usepackage{mathptmx}
\usepackage{booktabs}
\usepackage{graphicx}
\usepackage{bm}
\usepackage{amssymb}
\usepackage{amsmath}
\usepackage{ragged2e}
\usepackage[dvipsnames]{xcolor}
\definecolor{hgreen}{rgb}{0,.3,0}
\definecolor{hred}{rgb}{.3,0,0}
\definecolor{hblue}{rgb}{0,0,.3}
\definecolor{LightGray}{gray}{0.95}
\usepackage[
	colorlinks=true,
	linkcolor=hblue,
	citecolor=hgreen,
	filecolor=hblue,
	urlcolor=hred]{hyperref}

\newcommand{\gsim}{\gtrsim}
\newcommand{\lsim}{\lesssim}

\newcommand{\lc}{\left[}
\newcommand{\rc}{\right]}
\newcommand{\lp}{\left(}
\newcommand{\rp}{\right)}
\newcommand{\eps}{\varepsilon}

\newcommand{\mcO}{\mathcal{O}}

\newcommand{\alpg}{\texttt{ALPGEN}~}

\newcommand{\cv}{c_V}
\newcommand{\cvv}{c_{2V}}
\newcommand{\ccc}{c_3}
\newcommand{\dcvv}{\delta_{\cvv}}
\newcommand{\dccc}{\delta_{\ccc}}
\newcommand{\pT}[1]{p_{T_{#1}}}
\newcommand{\xtento}[1]{\times 10^{#1}}

\usepackage{multirow}

\begin{document} 

\title{Higgs pair production in vector-boson fusion at the LHC and beyond}

\author{
Fady Bishara\thanksref{e1,addr1}
\and Roberto Contino\thanksref{e2,addr2,addr3,addr4}
\and Juan Rojo\thanksref{e3,addr5,addr6}
}

\institute{
Rudolf Peierls Centre for Theoretical Physics, 1 Keble Road, University of Oxford, UK\label{addr1}
\and Scuola Normale Superiore, Pisa and INFN Pisa, Italy\label{addr2}
\and Institut de Th\'eorie des Ph\'enomenes Physiques, EPFL, Lausanne, Switzerland\label{addr3}
\and Theoretical Physics Department, CERN, Geneva, Switzerland\label{addr4}
\and Department of Physics and Astronomy, VU University Amsterdam, De Boelelaan 1081, NL-1081, HV Amsterdam, The Netherlands\label{addr5}
\and Nikhef, Science Park 105, NL-1098 XG Amsterdam, The Netherlands\label{addr6}
}

\thankstext{e1}{\email{fady.bishara@physics.ox.ac.uk}}
\thankstext{e2}{\email{roberto.contino@sns.it}}
\thankstext{e3}{\email{j.rojo@vu.nl}}

\maketitle

\begin{abstract}
The production of  pairs of Higgs bosons at hadron colliders provides unique
information on the Higgs sector and on the mechanism underlying electroweak
symmetry breaking (EWSB). Most studies have concentrated on the gluon fusion
production mode which has the largest cross section. However, despite its small
production rate, the vector-boson fusion channel can also be relevant since even
small modifications of the Higgs couplings to vector bosons induce a striking
increase of the cross section as a function of the invariant mass of the Higgs
boson pair. In this work, we exploit this unique signature to propose a strategy
to extract the $hhVV$ quartic coupling and provide model-independent constraints
on theories where EWSB is driven by new strong interactions. We take advantage
of the higher signal yield of the $b\bar b b\bar b$ final state and make
extensive use of jet substructure techniques to reconstruct signal events with a
boosted topology, characteristic of large partonic energies, where each Higgs
boson decays to a single collimated jet . Our results demonstrate that the
$hhVV$ coupling can be measured with 45\% (20\%) precision at the LHC for
$\mathcal{L}=300$ (3000) fb$^{-1}$, while a 1\% precision can be achieved at a
100 TeV collider.
\end{abstract}

\input{sec-intro}

\input{sec-theory}
\input{sec-analysis}

\input{sec-results}
\input{sec-conclusions}

\appendix
\input{sec-mc}
\input{sec-fitting}
\input{sec-qcdmultijet}

\input{sec-fit-coeffs}

\bibliography{HHvbf}

\end{document}

%% file: sec-intro.tex
\section{Introduction}

Following the discovery of the Higgs boson in
2012~\cite{Aad:2012tfa,Chatrchyan:2012xdj}, the measurement of its couplings to
the other Standard Model (SM) particles has become one of the main goals of the
LHC programme. In this respect, double Higgs production provides a unique
handle, in particular since it allows the extraction of the trilinear Higgs
self-coupling $\lambda$. In addition to constraining $\lambda$, in the
vector-boson fusion (VBF) channel double Higgs production also probes the
strength of the  Higgs non-linear interactions with vector bosons at high
energies. This process can thus help establish the nature of the Higgs boson,
whether it is a composite or elementary state, and whether or not it emerges as
a Nambu-Goldstone boson (NGB) of some new dynamics at the TeV
scale~\cite{Giudice:2007fh,Contino:2010mh,Contino:2013gna}.

Many scenarios of new physics  Beyond the SM (BSM) generically predict enhanced
cross sections for Higgs pair production with or without the resonant production
of new intermediate states, see for example
Refs.~\cite{Belyaev:1999kk,Grinstein:2007iv,Grober:2010yv,Cao:2013si,Gouzevitch:2013qca,Barbieri:2013hxa,Ellwanger:2013ova,Han:2013sga,No:2013wsa,Hespel:2014sla,Cao:2014kya,Azatov:2015oxa,Buttazzo:2015bka,Wen-Juan:2015gqg,Wu:2015nba}. For this reason, searches for Higgs pair production at the LHC by ATLAS and CMS have already started at $8\,$TeV~\cite{Aad:2014yja,Khachatryan:2015yea,Aad:2015uka,Aad:2015xja,Khachatryan:2015tha} and $13\,$TeV~\cite{ATLAS:2016qmt,Aaboud:2016xco,CMS:2016pwo,CMS:2016cdj}, and will continue during Runs II and III, as well as at the High-Luminosity LHC (HL-LHC) upgrade with $3\,\text{ab}^{-1}$ of integrated luminosity. On the other hand, in the SM the small production rates make a measurement of Higgs pair production extremely challenging even at the HL-LHC, and the ultimate accuracy could only be achieved at a future $100\,$TeV hadron collider~\cite{Yao:2013ika,Barr:2014sga,Papaefstathiou:2015iba,Arkani-Hamed:2015vfh,Contino:2016spe,Azatov:2015oxa}.

Similarly to single Higgs production, the dominant
mechanism for Higgs pair production is the gluon-fusion
mode~\cite{deFlorian:2016spz}. This channel has been extensively studied in the
literature and several  final states have been considered, including $b\bar
b\gamma\gamma$, $b\bar{b}\tau^+\tau^-$, $b\bar{b}W^+W^-$ and $b\bar{b}b\bar{b}$
(for a list of feasibility studies, see for example
Refs.~\cite{Dolan:2012rv,Papaefstathiou:2012qe,Barr:2013tda,Cooper:2013kia,Yao:2013ika,Lu:2015jza,Wardrope:2014kya,deLima:2014dta,Behr:2015oqq,Azatov:2015oxa}).
Working in the infinite top mass approximation, the gluon-fusion di-Higgs production cross section was calculated at NLO in~\cite{Dawson:1998py} and NNLO in~\cite{deFlorian:2013jea}. The resummation of soft-gluon emissions was performed at NNLL in~\cite{Shao:2013bz,deFlorian:2015moa}. Beyond the $m_t\to\infty$ limit, the impact of top quark mass effects on NLO QCD corrections was first determined in~\cite{Maltoni:2014eza} through a reweighting technique based on an approximate two-loop matrix element and by~\cite{Grigo:2013rya,Degrassi:2016vss} in a $1/m_t$ expansion.
Recently, the full NLO calculation was performed by~\cite{Borowka:2016ehy}.
Matching the fixed order computations to a parton shower was done at LO in~\cite{Maierhofer:2013sha} and at NLO in~\cite{Frederix:2014hta}.


Recent studies indicate that Higgs pair production in gluon-fusion at the HL-LHC will allow the extraction of the Higgs self-coupling $\lambda$ with $\mathcal{O}(1)$ accuracy, with details varying with the analysis and the specific final state, see Refs.~\cite{ATL-PHYS-PUB-2014-019,ATL-PHYS-PUB-2015-046,CMS:2015nat,ATL-PHYS-PUB-2016-024,CMS-DP-2016-064} for the latest ATLAS and CMS estimates, as well as~\cite{Barr:2013tda,Goertz:2013kp,Yao:2013ika,Slawinska:2014vpa,Azatov:2015oxa}.

Higgs pairs can also be produced in the  VBF
channel~\cite{Giudice:2007fh,Contino:2010mh,Dolan:2013rja,Brooijmans:2014eja,Liu-Sheng:2014gxa,Dolan:2015zja}
where a soft emission of two vector bosons from the incoming protons is followed by 
the hard $VV \to hh$ scattering, with $V=W,Z$.
In the SM, the VBF inclusive cross section
at 14 TeV is around 2 fb -- more than one
order of magnitude smaller than in gluon fusion.
QCD corrections give a 10\% increase and have been computed at NLO in Refs.~\cite{Baglio:2012np,Frederix:2014hta} and at NNLO in Ref.~\cite{Liu-Sheng:2014gxa}.
Production in association with $W$ or $Z$ bosons, known as the Higgsstrahlung process~\cite{Barger:1988jk,Baglio:2012np,Cao:2015oxx},
or with top quark pairs~\cite{Englert:2014uqa}, exhibit even smaller cross sections.

Despite its small rate, Higgs pair production via VBF is quite interesting since
even small modifications of the SM couplings can induce a striking increase of the
cross section as a function of the di-Higgs mass.
Specific models leading to this behaviour are, for instance, those
where the Higgs is a composite pseudo-NGB (pNGB) of new strong dynamics at the TeV scale~\cite{Kaplan:1983fs}.
In these theories, the Higgs anomalous couplings imply a growth of the $VV\to hh$ cross section with the
partonic center-of-mass energy, $\hat{\sigma} \propto \hat s/f^4$, where $f$ is the pNGB decay constant~\cite{Giudice:2007fh}.
This enhanced sensitivity to the underlying strength of the Higgs interactions makes 
double Higgs production via VBF
a key process to test the nature of the electroweak symmetry breaking dynamics
and to constrain the $hhVV$ quartic coupling.
A first study of double Higgs production via VBF at the LHC 
was performed in Ref.~\cite{Contino:2010mh}, for a mass $m_{h}=180\,$GeV, by
focusing on the $4W$ final state.
Following the discovery of the Higgs boson,
more studies of the $hhjj$ process at the LHC were presented in Refs.~\cite{Dolan:2013rja,Brooijmans:2014eja,Dolan:2015zja,Nakamura:2016agl}.

In this work, we revisit the feasibility of VBF Higgs pair production at the LHC
and focus on the $hh\to b\bar{b}b\bar{b}$ final state. While this final state
benefits from increased signal yields due to the large branching fraction of
Higgs bosons to bottom quarks, ${\rm BR}\!\lp H\to b\bar{b}\rp=0.582$ in the
SM~\cite{deFlorian:2016spz}, it also suffers from overwhelming
large QCD multi-jet backgrounds. In this respect, the remarkable VBF topology,
characterized by two forward jets well separated in rapidity and with a large
invariant mass, together with a reduced hadronic activity in the central region,
provides an essential handle to disentangle signal events  from the QCD
background. Additionally, the di-Higgs system will acquire a substantial boost
in the presence of BSM dynamics. It is thus advantageous to resort to jet
substructure techniques~\cite{Salam:2009jx} in order to fully exploit the
high-energy limit and optimize the signal significance.

We will thus focus on the kinematic region where the invariant mass of the Higgs
pair, $m_{hh}$, is large because modifications of the  couplings between the
Higgs and vector bosons cause the tail of this distribution to become
harder in the signal whereas the background is not modified. Therefore, this
region exhibits the highest sensitivity to the modified Higgs couplings and
in particular to the deviations in  the $hhVV$ quartic coupling $\cvv$. Given
that for large $m_{hh}$ the Higgs bosons can be produced boosted, improved
discrimination can be achieved using jet substructure, and to this end we use
scale-invariant tagging~\cite{Gouzevitch:2013qca,Behr:2015oqq} to smoothly
combine the resolved, intermediate and boosted topologies.

Our analysis takes into account all the main reducible and irreducible
backgrounds: QCD multijet production, Higgs production via gluon fusion (where
additional radiation can mimic the VBF topology), and top quark pair production.
We pay special attention to the role of light and charm jets being misidentified
as $b$-jets which can contribute sizeably to the total background yield. For
instance, in the $gg\to hh\to b\bar{b}b\bar{b}$ channel,  the $2b2j$ background 
is comparable to the $4b$ component~\cite{Behr:2015oqq}.

We quantify the constraints on the Higgs quartic coupling $\cvv$ that can be
obtained from VBF di-Higgs production at the LHC 14 TeV  with
$\mathcal{L}=300\,\text{fb}^{-1}$ and  $3000\,\text{fb}^{-1}$ as well as at a
Future Circular Collider (FCC) with a centre-of-mass energy of 100 TeV and a
total luminosity of $10\,\text{ab}^{-1}$. We find that, despite the smallness of the
production cross sections, the LHC with $300\,\text{fb}^{-1}$ can already
constrain the $hhVV$ coupling with an accuracy of $_{-37\%}^{+45\%}$ around its
SM value at the 1-$\sigma$ level, which is further reduced to $_{-15\%}^{+19\%}$
at the HL-LHC and  down to the 1\% level at the FCC. Our results strongly
motivate that searches for VBF Higgs pair production at the LHC should already
start during Run II.

The structure of this paper is as follows. In Sec.~\ref{sec:theory} we present the
general parametrization of the Higgs couplings which we
adopt and review its impact
on VBF Higgs pair production.
Then in Sec.~\ref{sec:analysis}, we discuss the analysis
strategy used to disentangle the signal from the background events in the
$b\bar{b}b\bar{b}$ final state.
Our main results are presented in
Sec.~\ref{sec:results}, where we quantify the potential of the VBF di-Higgs process
to measure the $hhVV$ coupling at various colliders and
discuss the validity of the effective field theory expansion.
Finally in Sec.~\ref{sec:conclusions}, we conclude and discuss how
our analysis strategy  could be applied to 
related processes.
Technical details are collected in three appendices
which describe the Monte Carlo event generation
of signal and background events (\ref{sec:mc}),
the fits to the tail of the $m_{hh}$ distribution
for backgrounds (\ref{sec:fitting}), and the
validation studies of the QCD multijet event generation
(\ref{sec:qcdmultijet}).


%% file: sec-theory.tex
\section{Higgs pair production via vector boson fusion at hadron colliders}
\label{sec:theory}

We begin by reviewing the theoretical framework for Higgs pair production 
via vector boson fusion in hadronic collisions.
First, we introduce a general
parametrization of the Higgs couplings in the effective field theory (EFT)
framework.
Then, we turn to consider the values that these 
couplings take in specific models.
Finally, we briefly discuss the validity of the EFT approximation and the possible
contribution of
heavy resonances to this process.

\subsection{General parametrization of Higgs couplings}

Following Ref.~\cite{Contino:2010mh}, we introduce a general parametrization of the
couplings of a light Higgs-like scalar $h$ to the SM vector bosons and fermions.
At energies much lower than the mass scale of any new resonance, the theory is described by an
effective Lagrangian obtained by making a derivative expansion.
Under the request  of custodial symmetry, the three NGBs associated with electroweak symmetry breaking parametrize the coset $SO(4)/SO(3)$ 
and can be fitted into a $2\times 2$ matrix
\begin{equation}
\Sigma =e^{i\sigma^a\pi^a/v} \, ,
\end{equation}
with $v=246\,$GeV the Higgs vacuum expectation value.
Assuming that the couplings of the Higgs boson to SM fermions scale with their masses and
do not violate flavor, the resulting effective Lagrangian in~\cite{Contino:2010mh}
can be parametrized as
\begin{equation}
\label{eq:lagrangian}
\begin{split}
{\cal L}  \supset
 & \, \frac{1}{2}(\partial_\mu h)^2 - V(h) +\frac{v^2}{4}{\rm Tr}\!\left (D_\mu \Sigma^\dagger D^\mu \Sigma\right )
    \left [ 1+2c_V\, \frac{h}{v}+c_{2V}\,\frac{h^2}{v^2}+\dots\right ] \\
 &-m_i\,\bar \psi_{Li}\, \Sigma\left (1+c_{\psi}\,\frac{h}{v} + \dots \right)\psi_{Ri}\,+\,{\rm h.c.}\, ,
\end{split}
\end{equation}
where $V(h)$ denotes the Higgs potential, 
\begin{equation}
\label{eq:potential}
V(h) = \frac{1}{2} m_h^2 h^2 + c_3\, \frac{1}{6} \left( \frac{3m_h^2}{v} \right) h^3 + c_4\, \frac{1}{24} \left( \frac{3m_h^2}{v^2} \right) h^4 + \dots
\end{equation}
The parameters $\cv$, $\cvv$, $c_{\psi}$, $\ccc$, and $c_4$ are in general
arbitrary coefficients, normalized so that they equal 1 in the SM.
The Higgs mass is fixed to be
$m_h=125$ GeV~\cite{Aad:2015zhl}.

\begin{figure}[t]
\begin{center}
\includegraphics[width=\textwidth]{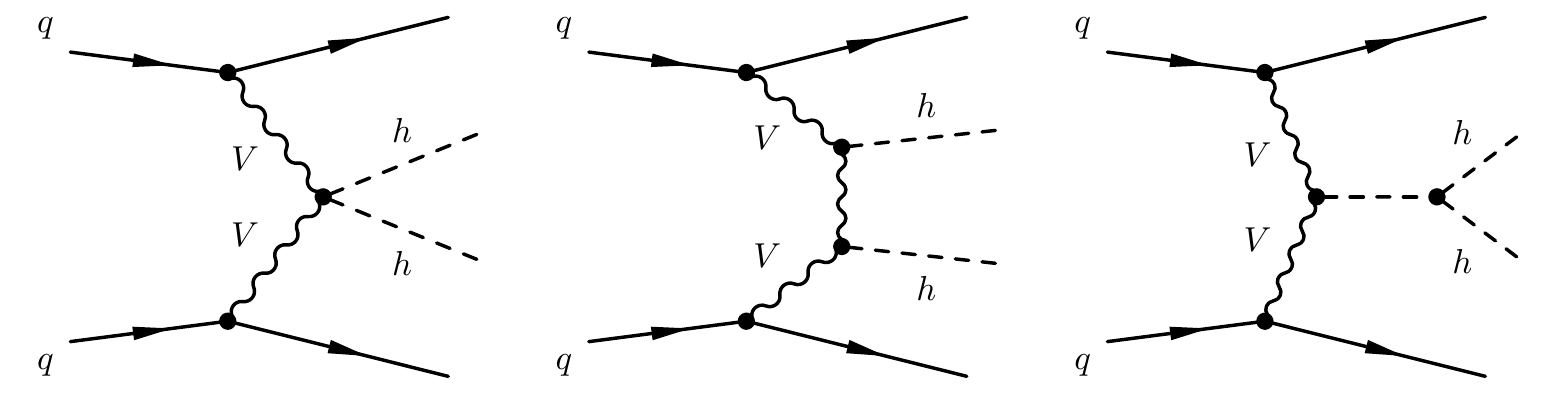}
\caption{\small 
Tree-level Feynman diagrams contributing to Higgs pair
production via VBF.
In terms of Eq.~(\ref{eq:lagrangian}), the left, middle,
and right diagrams scale with
$c_{2V}$, $c_V^2$, and $c_{V}c_3$, respectively.
 \label{fig:feyn}} 
\end{center}
\end{figure}

As the notation in Eq.~(\ref{eq:lagrangian}) indicates,
the coefficients $c_V$, $\cvv$, and $\ccc$ control the strength of the $hVV$,
$hhVV$ and $hhh$ couplings, respectively.
The coefficients $c_{\psi}$ and $c_4$ instead modify the Higgs coupling to fermions and quartic self interaction.
Thus, they do not affect the double-Higgs production cross section in the VBF channel.
In Fig.~\ref{fig:feyn},
we show the tree-level Feynman diagrams, in the unitary gauge, that contribute
to Higgs pair production  in the  vector-boson fusion channel at hadron colliders.
 In terms of the general parametrization of Eq.~(\ref{eq:lagrangian}), the left, middle,
and right diagrams scale with
$c_{2V}$, $c_V^2$, and $c_{V}c_3$, respectively.

In the SM, a cancellation dictated by perturbative unitarity occurs between the 
first and second diagrams.
This is best understood by describing the process as a slow emission of the vector bosons by the protons
followed by
their hard scattering into a pair of Higgs bosons~\cite{Chanowitz:1985hj}.
For generic values of $c_V$ and $c_{2V}$,
the amplitude of the partonic scattering $VV\to hh$ grows with the energy $\sqrt{\hat s}$
until the contribution
from the new states at the cutoff scale $\Lambda$ unitarizes it.
The leading contribution in the energy range $m_W \ll \sqrt{\hat s} \equiv m_{hh} \ll \Lambda$ comes from the scattering of longitudinal vector bosons and is given by
\begin{equation}
\label{eq:growth}
{\cal A}({V_L V_L\to hh}) \simeq  \frac{\hat{s}}{v^2}(c_{2V}-c_V^2)\, ,
\end{equation}
up to $\mathcal{O}(m_W^2/\hat s)$ and $\mathcal{O}(\hat s/\Lambda^2)$ corrections.
In scenarios with $c_{2V} \ne c_V^2$, the growth of the partonic cross section with $\hat s$
thus provides a smoking-gun signature for the presence of BSM dynamics~\cite{Giudice:2007fh}.

In the parametrization of Eq.~(\ref{eq:lagrangian}), the amplitude for the process $pp \to hh jj$ can be decomposed as follows
\begin{equation}
\mathcal{A} \, = \, \widetilde{A}\,c_V^2 + \widetilde{B}\,c_{2V} + \widetilde{C}\,c_Vc_3 \,,
\end{equation}
where $\widetilde{A}$, $\widetilde{B}$, and $\widetilde{C}$ are numerical
coefficients.
In the present work, we will focus on the quartic coupling
$\cvv$ and set $\cv$ and $\ccc$ to their SM values.
This is justified for $\cv$ since the ATLAS and CMS measurements of Higgs production cross sections, when analysed
in the context of a global fit of Higgs properties~\cite{Khachatryan:2014ira,Aad:2015tna,Aad:2015gba} typically set 
bounds on $\cv-1$ at the level of $10-20\%$, depending on the specific
assumptions made -- see for
example~\cite{Espinosa:2012im,Ellis:2013lra,Pomarol:2013zra} and references therein.
Tighter limits on $c_V$ can be derived from electroweak precision tests in the absence of additional BSM
contributions~\cite{Ciuchini:2013pca}.

On the other hand, the trilinear Higgs coupling~$c_3$ (where $c_3 = \lambda/\lambda_{\text{\sc sm}}$)
only has loose experimental constraints so far.
As an illustration, a recent ATLAS search for non-resonant
Higgs pair production at 13 TeV in the $b\bar{b}b\bar{b}$
final state~\cite{Aaboud:2016xco} translates into the bound
$\sigma(hh)/\sigma_\text{\sc sm}(hh)\lsim 27$ at
the 95\% confidence level.
Achieving $\mathcal{O}(1)$ precision in the measurement of $\ccc$ will thus most
likely require the full HL-LHC statistics. Focusing on VBF production, as
anticipated and further discussed in the following, gaining sensitivity to
$\cvv$ is achieved by reconstructing events with large values of $m_{hh}$. In
this kinematic region, it turns out that the sensitivity to $\ccc$ is reduced,
indicating that our analysis is not optimal to probe the Higgs trilinear
coupling.
For these reasons,
setting $c_V = c_3 =1$ is a good approximation in the context of the present analysis.
We can then define
\begin{equation}
\delta_{c_{2V}} \equiv \cvv-1\, ,
\label{eq:delta2}
\end{equation}
and this way the total cross section will be parametrized as
\begin{equation}
\label{eq:xsecgeneral}
	\sigma = \sigma_\text{\sc sm}\left(1+A\,\dcvv + B\,\dcvv^2\right)\, .
\end{equation}
However, while setting $\ccc=1$ is a very good approximation, fixing $\cv=1$ is not as equally well justified.
In particular, it would be more prudent to treat $\cv$ as a Gaussian distributed nuisance parameter centred around its SM value with a width corresponding to the current experimental precision.
To do this, a similar expression to Eq.~(\ref{eq:xsecgeneral}) above can be derived by neglecting the sub-leading effects involving $\ccc$.
In this case, Eq.~(\ref{eq:xsecgeneral}) is replaced by
\begin{equation}
	\sigma \approx \sigma_\text{\sc sm}\,\cv^4\left(1+A\,\left[\frac{\cvv}{\cv^2}-1\right] + B\,\left[\frac{\cvv}{\cv^2}-1\right]^2\right)\,.
\label{eq:xsecprime}
\end{equation}
We will use this expression to evaluate the impact of $\cv$ on the derived bounds on $\dcvv$ at the end of Sec.~\ref{sec:results}.
The values of the SM cross section $\sigma_\text{\sc sm}$ and of the parameters
$A$, $B$ are reported in Table~\ref{tab:ci-coeffs} for $\sqrt{s}=14$ and $100\,$TeV, both
after acceptance cuts and after applying all the analysis cuts as discussed in Sec.~\ref{sec:analysis}
-- see~\ref{sec:fit-coeffs} for the values of the parameters in bins of $m_{hh}$.
We will make extensive
use of this parametrization in Sec.~\ref{sec:results} where we  present our results in terms of the
sensitivity on $\delta_{\cvv}$.
Note that the value of $A$ and $B$ increase after imposing all cuts
precisely because they have been optimized to enhance the sensitivity on $c_{2V}$.

%
\begin{table}[t]\centering
	\renewcommand{\arraystretch}{1.3}
	\begin{tabular}{c c c c c c c}\toprule[1pt]
		$\sqrt{s}$ & Cuts & $\sigma_\text{\sc sm}$ [fb] &  $A$ & $B$ & $C$ & $D$\\\midrule
		\multirow{2}{*}{$14\,$TeV} & Acceptance & 0.010 &  $-5.19$ & 29.5 & $-0.939$ & 0.854 \\
		& All & 0.0018 & $-8.18$ & 67.5 & $-0.699$ & 0.325 \\
			\midrule
		\multirow{2}{*}{$100\,$TeV} & Acceptance & 0.20 & $-9.18$ & 306 & $-0.699$ & 0.584 \\
		& All & 0.030 & $-20.7$ & 1080 & $-0.516$ & 0.251 \\
			\bottomrule[1pt]
	\end{tabular}
	\caption{\small Coefficients of Eqs.~(\ref{eq:xsecgeneral}) and~(\ref{eq:xsec-dc3}) as obtained through a fit of Montecarlo points.
        The cuts are listed in Table~\ref{tab:sel-cuts}
	and Eqs.~(\ref{eq:higgsmasswindow})--(\ref{eq:mhhcut}).
	}
	\label{tab:ci-coeffs}
\end{table}
%

Although we do not attempt to extract $c_3$ with our analysis, it is still interesting to discuss the dependence of the total cross section
on this parameter.
By fixing $c_V = c_{2V} =1$ and defining $\delta_{\ccc}\equiv\ccc-1$, the cross section can now be parametrized as
\begin{equation}
\sigma = \sigma_\text{\sc sm}\left(1+C\,\delta_{\ccc} + D\,\delta_{\ccc}^2\right).
	\label{eq:xsec-dc3}
\end{equation}
The coefficients $C$ and $D$ are also reported in Table~\ref{tab:ci-coeffs}.
As opposed to the previous case, now
their values decrease after applying the full set of cuts, reflecting
that the sensitivity on $c_3$ is
suppressed by our analysis which aimed at measuring $c_{2V}$.
Extracting $\ccc$ would require retaining the events close to the $hh$ threshold, but this kinematic region is totally dominated by the background and turns out to be of little use.
A measurement of the Higgs trilinear coupling
in the VBF channel using the the $b\bar b b \bar b$ final state
thus does not seem feasible even at the FCC.
Though, other final states might exhibit better prospects at 100 TeV.

\subsection{Models}
\label{sec:scenarios}

The Lagrangian of Eq.~(\ref{eq:lagrangian}), with arbitrary values of
the coefficients $c_V$, $c_{2V}$, $c_{\psi}$, $c_3$ and $c_4$,
describes a generic 
light scalar, singlet of the custodial symmetry,
independently of its role in the electroweak symmetry breaking.
In specific
UV models, however, the coefficients $c_i$ are generally related
to each other
and their values are
subject to constraints which depend on whether the Higgs-like boson $h$
is part of an $SU(2)_L$ doublet.
For example, in the SM, all the parameters in Eq.~(\ref{eq:lagrangian}) are equal to 1 and terms denoted by the ellipses vanish.
In this case, the scalar~$h$ and the three NGBs combine to form a doublet of $SU(2)_L$ which is
realized linearly at high energies.

Composite Higgs theories are another example where the electroweak symmetry is realized linearly in the UV, though, in this case non-linearities
in the Higgs interactions can be large and are controlled by the ratio $\xi\equiv v^2/f^2$, where $f$ is the pNGB decay constant.
For instance, minimal $SO(5)/SO(4)$ models~\cite{Agashe:2004rs,Contino:2006qr} predict
\begin{equation}
c_V=\sqrt {1-\xi} \, ,\qquad c_{2V}=1-2\xi\, .
\label{abgoldstone}
\end{equation}
On the other hand, the value of the Higgs trilinear coupling is not
determined by the coset structure alone, and depends on how the Higgs potential is  specifically generated.
For instance, in the  MCHM5 model with fermions transforming as vector
representations of $SO(5)$~\cite{Contino:2006qr},
the Higgs potential is entirely generated by loops of SM fields and the Higgs
trilinear coupling is predicted to be
\begin{align}
c_3 =& \frac{1-2\xi}{\sqrt {1-\xi}}\, .
\label{eq:d3MCHM5}
\end{align}

A precision model-independent determination of $c_{2V}$ would thus provide
stringent constraints on a number of BSM scenarios.
To begin with, if the Higgs boson belongs to an electroweak doublet, as suggested by the LHC data, and the modifications to its couplings are small,
then the values of $c_{2V}$ and $c_{2V}$ are in general predicted to be correlated~\cite{Giudice:2007fh,Contino:2013gna}:
\begin{equation} \label{eq:dilaton}
\delta_{c_{2V}} \simeq 2\, \delta_{c_V^2}\, ,
\end{equation}
where $\delta_{c_V^2} \equiv c_V^2 -1$.
This follows because there is a single dimension-6 effective operator ($O_H$ in the basis of Ref.~\cite{Giudice:2007fh}) which controls the shift
in both couplings.
Therefore, a high-precision measurement
of $c_{2V}$ can test whether the Higgs boson belongs to a doublet in case
a deviation is observed in $c_V$~\cite{Contino:2013gna}.

Another interesting case is the
scenario where the Higgs-like boson is not part of a doublet, and in fact does
not play any role in the electroweak symmetry breaking mechanism,
known as the light 
dilaton scenario~\cite{Halyo:1991pc,Goldberger:2008zz,Vecchi:2010gj,Campbell:2011iw,Chacko:2012vm,Bellazzini:2012vz}.
In this model, invariance under dilatations implies $\delta_{c_{2V}} = \delta_{c_V^2}$, a condition that can be tested if the two couplings in Eq.~(\ref{eq:dilaton}) can be measured with
comparable precision.
Moreover, comparing $c_{2V}$ and $c_V$ can also provide
information on the coset structure in the case of 
a composite NGB Higgs~\cite{Contino:2013gna}.

A large variety of BSM scenarios also exist where
the Higgs trilinear coupling receives large modifications while the value
of the other couplings 
are close to the SM prediction.
Higgs portal models fall in this class, see for example the
discussion in~\cite{Dolan:2012rv,Azatov:2015oxa} and also~\cite{Craig:2014lda,Assamagan:2016azc}.
However, since our analysis is not sensitive to $c_3$, we will not consider these scenarios
any further and will always assume $\delta_{c_3}=0$.

In the following, we will take as a representative
benchmark scenario a model with $\cvv=0.8$
corresponding to $\dcvv=-0.2$, with the
other couplings set to their SM values,
namely $\delta_{c_V}=\delta_{c_3}=0$.

\subsection{Validity range of the effective theory}
\label{sect:partonresonances}

As discussed above, and shown by Eq.~(\ref{eq:growth}), the amplitude 
of the partonic scattering $VV\to hh$  grows with the energy in the EFT
described by Eq.~(\ref{eq:lagrangian}).
However, this behavior holds only below the typical mass scale of new BSM states, i.e. below the cutoff scale $\Lambda$ of the effective theory.
When the invariant mass of the di-Higgs
system becomes large enough, the EFT approximation breaks down
and it becomes necessary to take into full account  the contribution from the exchange of new
states, such as vector and scalar resonances.
These resonances eventually tame
the growth of the scattering amplitude at large energies
to be consistent with perturbative unitarity bounds.
In the context of  composite Higgs scenarios
the impact of resonances on $VV$ scattering has been
explored for instance in~\cite{Contino:2011np,Bellazzini:2012tv,Accomando:2012yg,Carcamo-Hernandez:2013ypa,Greco:2014aza,Pappadopulo:2014qza}.

While we do not include the effect of such resonances in this work, we
will report
the sensitivity on $\dcvv$ as a function of the maximum value of the invariant mass of the
di-Higgs system
 (see Fig.~\ref{fig:c2v-sens} in Sec.~\ref{sec:results}), 
as suggested by Ref.~\cite{Contino:2016jqw}.
This comparison
allows one to assess the validity of the EFT description once an estimate of
$c_{2V}$ is provided in terms of the 
masses and couplings of the UV dynamics.
The result of this analysis -- which is discussed in detail at the
end of Sec.~\ref{sec:results} --
confirms that the EFT is valid over the full range of $m_{hh}$ that is used
to derive limits on $\dcvv$.
The explicit inclusion of scalar and vector resonances and their
phenomenological implications for Higgs pair production via VBF
is left for future work.

%% file: sec-analysis.tex
\section{Analysis strategy}
\label{sec:analysis}

In this section we present our  analysis of double  Higgs production via VBF.
First, we discuss how signal and background events are reconstructed and classified.
This includes a description of  the jet reconstruction techniques
adopted, the $b$-tagging strategy, and
the event categorisation in terms of jet substructure.
Then we illustrate the various selection cuts imposed to maximize the signal significance,
in particular the VBF cuts,
as well as the method used to identify
the Higgs boson candidates.
Finally, we present the signal and background event rates for the various
steps of the analysis, and discuss how the signal cross-sections
are modified when  $c_{2V}$ is varied as compared to its SM value.

\subsection{Event reconstruction and classification}

Signal and background events are simulated at leading-order (LO) by means of
matrix-element  generators and then processed through a parton shower (PS).
The detailed description of the event generation of signal and backgrounds
can be found in~\ref{sec:mc}.
The dominant background is given by QCD multijet production,
while other backgrounds, such as top-quark pair production and Higgs pair production via
gluon-fusion, are much smaller.
After the parton shower, events are clustered with 
{\sc\small FastJet} v3.0.1~\cite{Cacciari:2011ma} using the
anti-$k_t$ algorithm~\cite{Cacciari:2008gp} with a jet radius $R=0.4$.

The resulting jets are processed through a $b$-tagging algorithm,
where a jet is tagged as $b$-jet with probability $\eps(b\text{-tag})$ 
if it contains a $b$-quark with $p_T^b > 15\,$GeV.
In order to account for $b$-jet misidentification (fakes),
jets which do not meet this requirement
are also tagged as $b$-jets with probability $\eps(c\text{-mistag})$ or $\eps(q,g\text{-mistag})$
depending on whether they contain a $c$-quark or not.
Only events with four or more jets, of which at least two
must be $b$-tagged, are retained at this stage.

Fully exploiting the $b\bar{b}b\bar{b}$ final state requires efficient
$b$-tagging capabilities, in both the resolved and boosted regimes as well as a
good rejection of fakes.
Both ATLAS and CMS  have presented recent studies of their capabilities
in terms of $b$-tagging and light-jet fake rejection for both topologies, see 
Refs.~\cite{CMS-DP-2013-005,CMS-PAS-BTV-13-001,ATL-PHYS-PUB-2014-013,ATL-PHYS-PUB-2014-014,CMS-DP-2015-038}
and references therein.
In the present study, we have
considered two representative $b$-tagging working points:
\begin{equation}
\begin{split}
	\text{WP1}&:  \hspace{0.4cm}
	\varepsilon(b\textrm{-tag})=0.75\,, \quad \varepsilon(c\textrm{-mistag}) = 0.1\,, \phantom{0}\quad \varepsilon(q,g\textrm{-mistag}) = 0.01\, , \\
	\text{WP2}&: \hspace{0.4cm}
	\varepsilon(b\textrm{-tag})=0.8\, ,\phantom{0} \quad \varepsilon(c\textrm{-mistag}) = 0.05\, , \quad \varepsilon(q,g\textrm{-mistag}) = 0.005\, .
\end{split}
\label{eq:btaggingWP}
\end{equation}
The first point is consistent with the current performance of ATLAS and CMS,
and is the one adopted as baseline in this paper.
The second working point is more
optimistic and is intended to assess how much one could gain with a more efficient $b$-tagger.
As our results will show (see Fig.~\ref{fig:significance}), 
using WP2 leads to a marginal improvement  in our analysis.
For simplicity, we applied the efficiencies in Eq.~\ref{eq:btaggingWP} to a jet
based on its constituents. This is sufficient for the purpose of the current
analysis which is, namely, to demonstrate the sensitivity of double Higgs production via VBF
to $\dcvv$. Accordingly, we leave a detailed study of $b$-tagging including
hadronization effects and $p_T$ dependence to future studies.

Subsequently to $b$-tagging, events are
classified through a scale-invariant tagging procedure~\cite{Gouzevitch:2013qca,Behr:2015oqq}.
This step is crucial to efficiently reconstruct the Higgs boson candidates and
suppress the otherwise overwhelming QCD backgrounds while at the same time
taking into account all the relevant final-state topologies.
The basic idea of this method is to robustly merge three event
topologies -- {\it boosted, intermediate} and {\it resolved} --
into a common analysis.
This is
particularly relevant for our study since,
as discussed in Sec.~\ref{sec:theory}, the degree of boost of the di-Higgs
system strongly depends
on the deviations of $\cvv$ from its SM value.

\begin{figure}[t]
  \centering
	\includegraphics[scale=1]{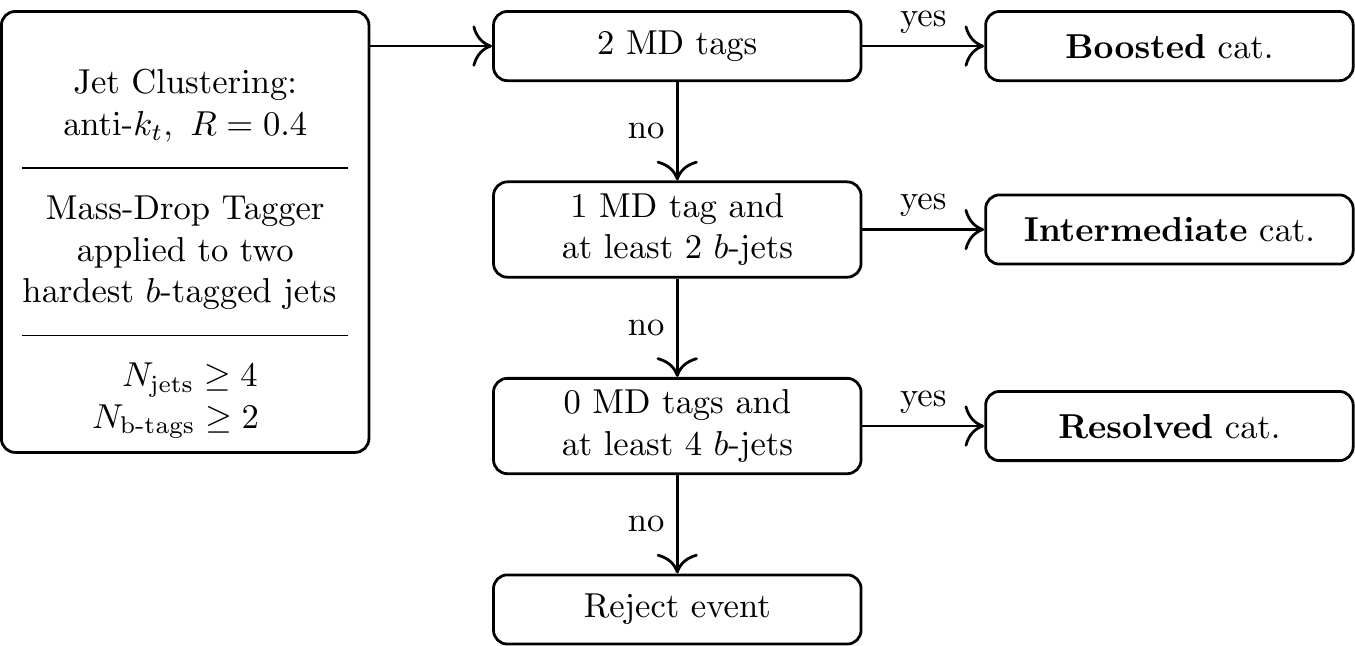}
 \caption{\small \label{fig:4b2j_cutflow} 
Schematic representation of the analysis strategy adopted in this work.
 }
\end{figure}

This scale-invariant tagging strategy is
schematically represented in Fig.~\ref{fig:4b2j_cutflow}.
First of all, the
$b$-tagged jets are ordered in $p_T$ and the constituents of the hardest two jets are then
re-clustered using the
Cambridge/Aachen (C/A) algorithm~\cite{Dokshitzer:1997in} with $R_{\rm C/A}=1.2$.
Each C/A jet is processed with the BDRS mass-drop (MD)
tagger~\cite{Butterworth:2008iy}.
This jet-substructure tagger has two
parameters: $\mu$ and $y_{\rm cut}$. And, in this work, we set $\mu = 0.67 $ and $y_{\rm cut} = 0.09$ as in the original BDRS study.
To determine if a
given jet arises from the decay of
a massive object, the last step of the clustering for
jet $j$ is undone, giving two subjets $j_1$ and $j_2$ which are ordered such that
$m_{j_1} > m_{j_2}$.
Then,  if the two subjets satisfy the conditions
\begin{equation}
m_{j_1}\le\mu\cdot m_j\qquad\text{and}\qquad
	\min(p_{Tj_1}^2,p_{Tj_2}^2)\,\Delta R^2_{j_1,j_2} > y_{\rm cut}\cdot m_{j}^2 \, ,
\label{eq:boosted-asymmetry-cut}
\end{equation}
where $\Delta R_{j_1,j_2}$ is the angular separation between the two subjets,
$j$ is tagged as a jet with a mass drop. Else, the procedure is applied recursively to
$j_1$ until a mass drop is found
or the C/A jet is fully unclustered.

Jets are mass-drop tagged only if they satisfy the following additional
requirement: at least two $b$-quarks must be contained within the jet, each of
which with $p_{Tb}\ge 15\,$GeV, and with a minimal angular separation $\Delta
R_{bb}\ge 0.1$. The request of a second $b$-quark completes our $b$-tagging
algorithm in the case of boosted jets. Other more sophisticated approaches to
$b$-tagging could have been considered -- e.g., using ghost-association between
large-$R$ MD-tagged jets and small-$R$ $b$-tagged
jets~\cite{Behr:2015oqq,Aad:2015uka}, or accounting for an efficiency which
depends on the jet $p_T$~\cite{Huffman:2016wjk}. The approach followed here is
at the same time simple  yet realistic enough for a first feasibility study with
the caveat that a more complete analysis should treat $b$-tagging more in line
with the actual performance of the ATLAS and CMS detectors (and in particular it
should include a full detector simulation).

The use of the  BDRS mass-drop tagger allows us to classify a given signal or
background event under one of the
three categories: {boosted}, if  two mass-drop tags are present; 
{ intermediate}, for an event with a single mass-drop tag; and
{ resolved}, if the event has no mass-drop tags.
In the resolved category,
events are only retained if they
contain at least 4 $b$-tagged jets, while  at least
2 $b$-tagged jets, in addition to the MD-tagged jet,
are required in the intermediate one.
By construction, this classification is exclusive, {\it i.e.}, each event is unambiguously
assigned to one of the three categories.
This exclusivity allows the consistent combination of the signal significance from the
three separate categories. 

Following the event categorisation,
acceptance cuts to match detector coverage are applied to signal and background events.
These cuts are listed in the upper part of Table~\ref{tab:sel-cuts}, and have
been separately optimized for the LHC $14\,$TeV and the FCC $100\,$TeV.
We require the $p_T$ of the light  ($b$-tagged) jets
to be larger than 25 GeV (25 GeV) at 14 TeV and than 40 GeV (35 GeV) at 100 TeV,
respectively.
Concerning the pseudo-rapidities of light
and $b$-tagged jets, $\eta_j$
and $\eta_b$, at the LHC the former is limited
by the coverage of the forward calorimeters,
while the latter is constrained
by the tracking region where $b$-tagging can be applied.
At 100 TeV, we assume
a  detector with extended coverage of the forward region up to $|\eta|$ of 6.5~\cite{SelvaggiM}.

\begin{table}[t]\centering\renewcommand{\arraystretch}{1.1}
	\begin{tabular}{lr @{\hskip 3mm} c @{\hskip 0.7cm} c }
		\toprule[1pt]
	&	&	14 TeV	&	100 TeV\\\midrule
\multirow{4}{*}{Acceptance cuts}	&	$p_{T_j}~\text{(GeV)}~\geq~$&	25	&	40\\
	&	$p_{T_b}~\text{(GeV)}~\geq~$&	25	&	35\\
	&	$|\eta_j|\leq~$				&	4.5	&	6.5\\
	&	$|\eta_b|\leq~$				&	2.5	&	3.0\\\midrule
\multirow{3}{*}{VBF cuts}	&	$|\Delta y_{jj}|\geq~$		&	5.0	&	5.0\\
&	$m_{jj}~\text{(GeV)}~\geq~$	&	700	&	1000\\
           &	Central jet veto: \ \ $p_{T_{j_3}}~\text{(GeV)}~\leq~$ 
		& 45 & 65\\
		\bottomrule[1pt]
	\end{tabular}
	\caption{\small Acceptance and VBF selection cuts applied to
          signal and background events after  jet clustering and $b$-tagging.
          The central jet veto is applied on jets with
          pseudo-rapidity $\eta_{j_3}$ in the interval $\eta_j^{\min} < \eta_{j_3} < \eta_j^{\max}$,
          where $\eta_j^{\max}$ and $\eta_j^{\min}$ are the
          pseudo-rapidities of the VBF-tagging jets.
        }
	\label{tab:sel-cuts}
\end{table}

\subsection{VBF selection cuts}
\label{sec:jets}

Subsequently to the acceptance cuts, we impose a set of selection cuts tailored
to the VBF topology which is characterized by two forward and very energetic
jets with little hadronic activity between them. In particular, we cut on the
rapidity separation $\Delta y_{jj}\equiv|y_j^{\rm lead}-y_j^{\rm sublead}|$ and
the invariant mass $m_{jj}$ of the two VBF-tagging jets, and impose a central
jet veto (CJV) on  the hardest non-VBF light jet in the central region. The VBF
tagging jets are defined as the pair of light jets satisfying the acceptance
cuts of Table~\ref{tab:sel-cuts} with the largest invariant mass $m_{jj}$. This
definition is robust with respect to soft contamination from the underlying
event (UE) and pile-up (PU) and to the contribution of $b$-jets mistagged as
light jets.

\begin{figure}[t]
	\centering
	\begin{minipage}{0.49\textwidth}\centering
		\hspace{1cm}LHC $14\,$TeV
		\includegraphics[width=\textwidth]{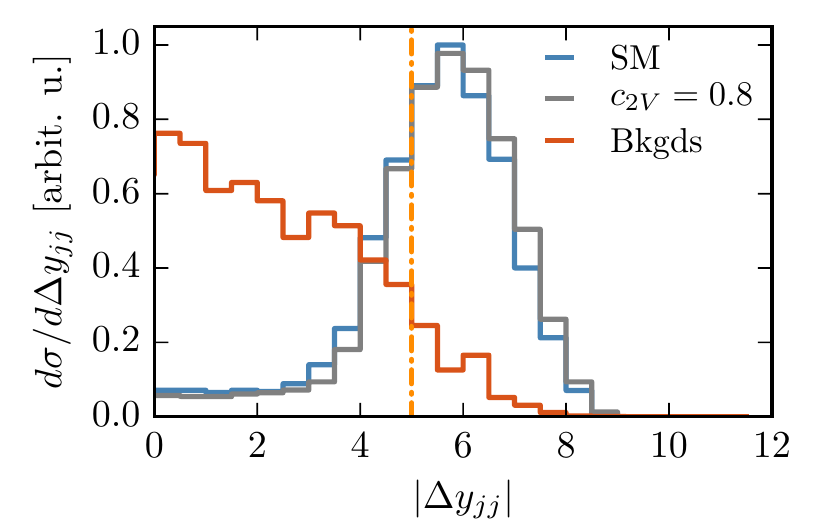}
		\includegraphics[width=\textwidth]{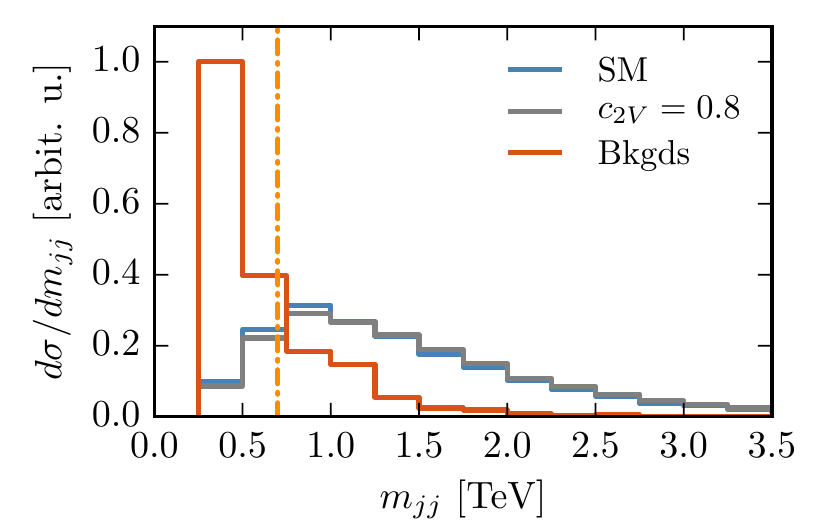}
	\end{minipage}
	\begin{minipage}{0.49\textwidth}\centering
		\hspace{1cm}FCC $100\,$TeV
		\includegraphics[width=\textwidth]{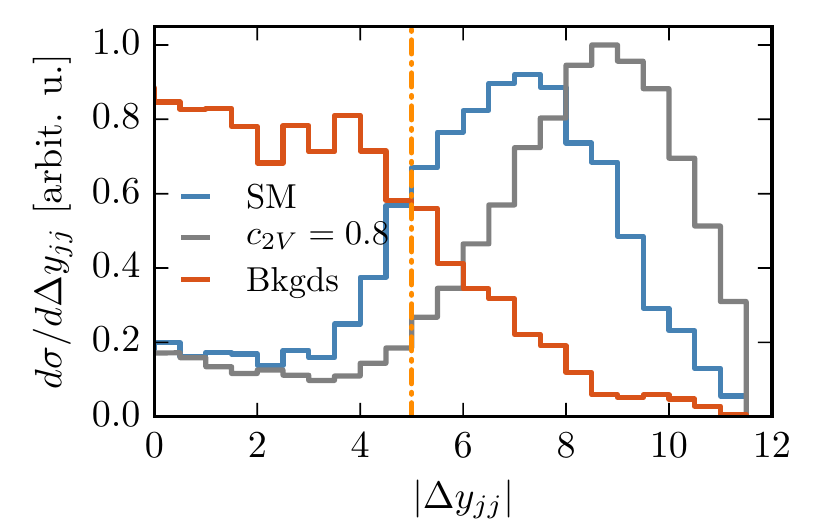}
		\includegraphics[width=\textwidth]{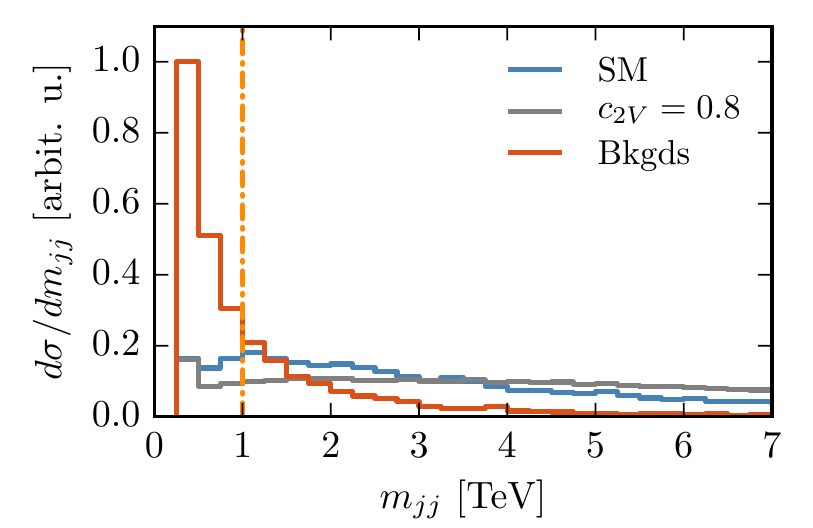}
	\end{minipage}
	\caption{\small \label{fig:dyljets} 
	Distribution of the rapidity separation $|\Delta y_{jj}|$ (upper) and
	the invariant mass $m_{jj}$ (bottom panels) of the VBF tagging
	jets at 14 TeV (left) and 100 TeV (right panels), for signal (SM and $c_{2V}=0.8$)
       and background events after the acceptance cuts.
        The 
	vertical line indicates the value of the corresponding cut from
	Table~\ref{tab:sel-cuts}.
        The distributions have been area-normalized and rescaled by a
	common factor.
	}
\end{figure}

Figure~\ref{fig:dyljets}  shows the distribution of the rapidity separation
$|\Delta y_{jj}|$ and invariant mass $m_{jj}$ of the VBF tagging jets at
$14\,$TeV  and $100\,$TeV after the acceptance cuts. In each case, we show the
results for the signal (SM and $c_{2V}=0.8$ benchmark) and for the total
background. The signal distributions exhibit the distinctive VBF topology, with
two VBF tagging jets widely separated in rapidity and with a large invariant
mass. This is in contrast with the backgrounds where both the $\Delta y_{jj}$
and $m_{jj}$ distributions peak at zero. In Fig.~\ref{fig:dyljets}, as well as
in the subsequent figures, kinematic distributions have been  area-normalized
and then rescaled by a common factor such that the largest bin in the plot
is of unit height.

Based on the distributions of Fig.~\ref{fig:dyljets} we identified appropriate
values of the VBF cuts, listed 
in Table~\ref{tab:sel-cuts} and represented in each panel by a vertical dash-dotted line.
It is important to tailor these cuts to the specific center-of-mass energy, 14 TeV and 100 TeV,
to avoid losing a substantial 
fraction of the signal events.
One should also take into account that the large rapidity separation between the VBF tagging jets
in signal events results from jets pairs with a large invariant mass, given that 
these two variables are strongly correlated~\cite{Contino:2010mh}.
This large separation in rapidity is especially useful in the $b\bar{b}b\bar{b}$ final state 
to trigger on signal events, providing
a significant improvement compared to
the same final state produced in gluon-fusion where triggering issues are more severe~\cite{deLima:2014dta,Behr:2015oqq}.

\begin{figure}[t]
	\centering
	\includegraphics[scale=1]{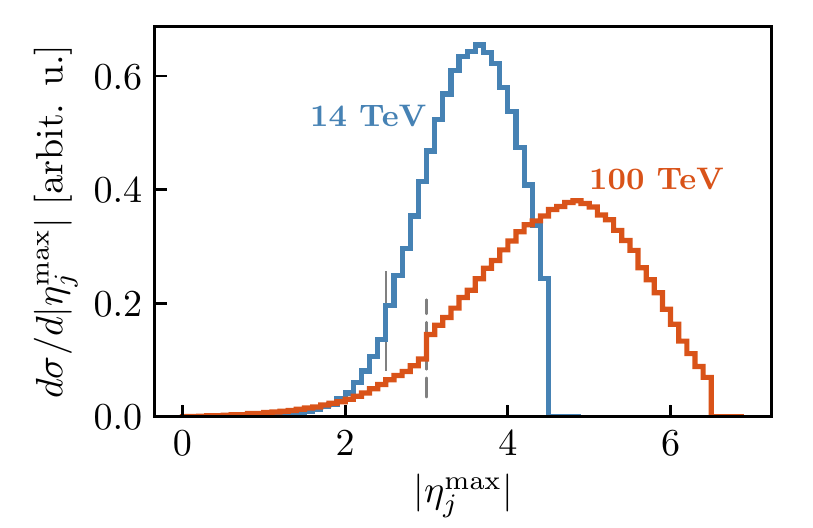}
        \caption{\small Distribution of the pseudo-rapidity $|\eta_j^{\rm max}|$ of the most forward light
	  jet at $14\,$TeV and $100\,$TeV.
	Both curves have a discontinuity at the edge of the corresponding b-tagging region which is delineated by the solid (dashed) grey vertical lines in the case of 14 (100) TeV. This discontinuity is more clearly visible in the 100 TeV curve and is purely due to combinatorics.}
\label{fig:dyljets2} 
\end{figure}

Figure~\ref{fig:dyljets} clearly highlights that in order
to maximize the acceptance of events with VBF topology -- the detectors must have a good
coverage of the forward region.
This issue is particularly relevant at $100\,$TeV as
illustrated in Fig.~\ref{fig:dyljets2} which shows the pseudo-rapidity distribution of the
most forward light jet.
%
At $100\,$TeV, this peaks at around~$5$, so a detector instrumented only
up to $|\eta|=4.5$ would lose  more than 50\% of  signal events.
 The discontinuity in the FCC
 case delineates the edge of the $b$-jet acceptance region, $|\eta|\le 3$,
 above which no $b$-tagging is attempted and $b$-jets contribute
to the light-jet yield.

\begin{figure}[t]
	\centering
	\begin{minipage}{0.49\textwidth}\centering
		\hspace{1cm}LHC $14\,$TeV
		\includegraphics[width=\textwidth]{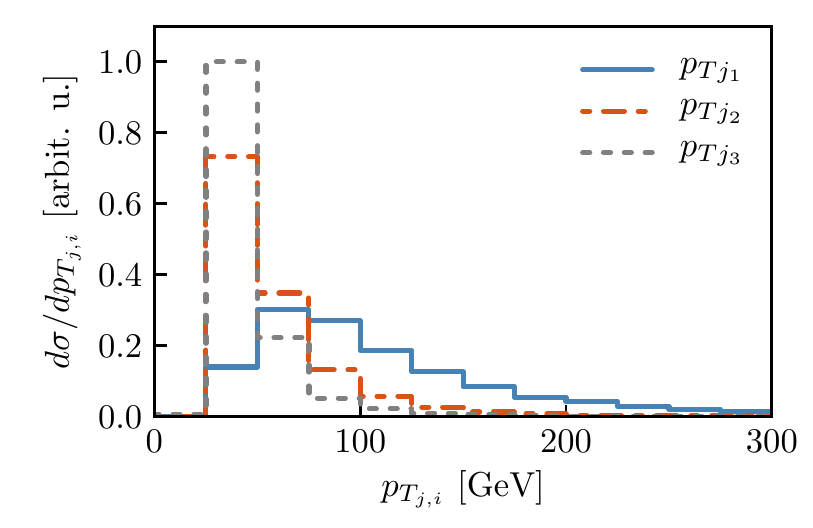}
	\end{minipage}
	\begin{minipage}{0.49\textwidth}\centering
		\hspace{1cm}FCC $100\,$TeV
		\includegraphics[width=\textwidth]{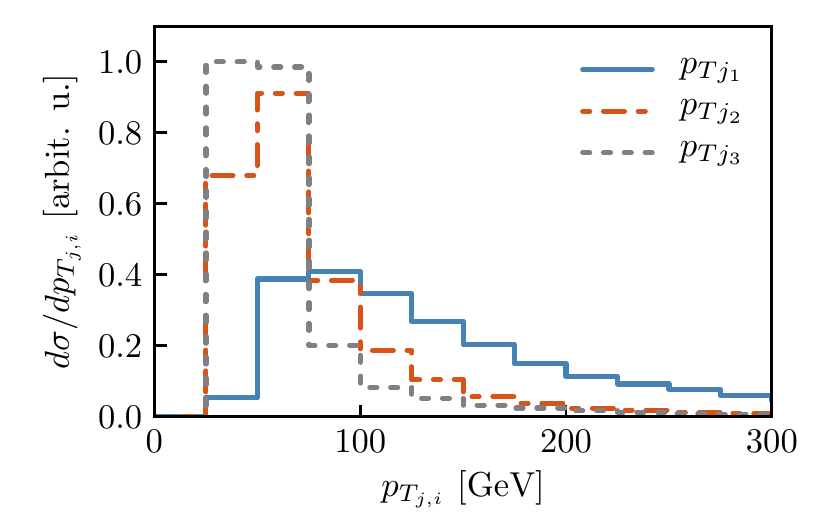}
	\end{minipage}
	\caption{\small \label{fig:ptljets} 
		Distributions of the $p_T$ of the leading, subleading, and 3\textsuperscript{rd}
		light jet at $14\,$TeV (left panel) and $100\,$TeV (right panel) for the SM signal.
		Distributions are area-normalized as in Fig.~\ref{fig:dyljets}.
	}
\end{figure}

Turning to the transverse momentum of the light jets, Fig.~\ref{fig:ptljets}
shows the $p_T$ distributions of the three hardest jets at 14 TeV and $100\,$TeV
for SM signal events. One can see that while the leading jet is typically quite
hard, the subleading ones are rather soft. It is thus important to avoid
imposing a too stringent cut in $\pT{}$, in order not to suppress the signal.
Fortunately, in contrast from the gluon-fusion process, adopting a soft $\pT{}$
cut is not a problem since triggering can be performed based on the VBF
topology. Comparing the $\pT{}$ distributions at $14\,$TeV and $100\,$TeV, their
shapes turn out to be rather similar, shifted towards larger values at
$100\,$TeV. This justifies the harder $\pT{}$ cut in this case (see
Table~\ref{tab:sel-cuts}), also required to reduce the contamination from UE and
PU.

As mentioned above, another characteristic feature of VBF production is a
reduced hadronic activity in the central region between the two VBF-tagging 
jets. This follows because the latter are not colour-connected since the
production of the central system only involves electroweak bosons. For this
reason, a CJV cut  is commonly imposed  in VBF analyses. This cut vetoes light
jets, with pseudo-rapidity $\eta_{j_3}$, lying between those of the VBF-tagging
jets, $\eta_j^{\max}>\eta_{j_3}>\eta_j^{\min}$, above a given $\pT{}$ threshold.

\begin{figure}[t]
	\centering
	\begin{minipage}{0.49\textwidth}\centering
		\hspace{1cm}LHC $14\,$TeV
		\includegraphics[width=\textwidth]{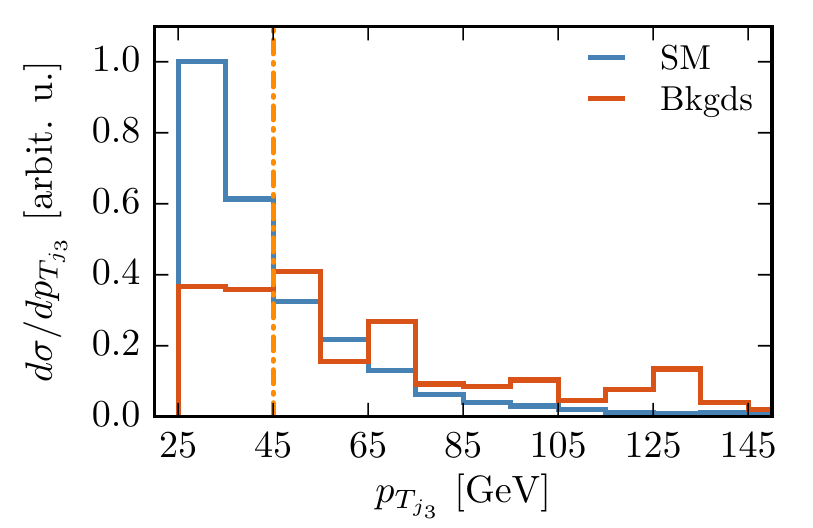}
	\end{minipage}
	\begin{minipage}{0.49\textwidth}\centering
		\hspace{1cm}FCC $100\,$TeV
		\includegraphics[width=\textwidth]{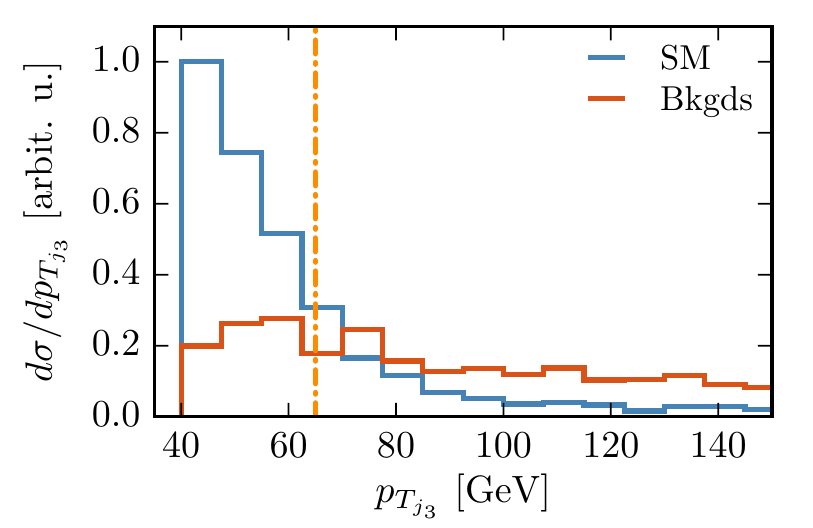}
	\end{minipage}
	\caption{\small \label{fig:cenjetveto} 
		Distribution of the $p_T$ of the 3\textsuperscript{rd}
		light jet at $14\,$TeV
		(left panel) and $100\,$TeV (right panel) for the SM signal and the
                total background, including only
                events  where this jet lies within
		the pseudo-rapidity region between the VBF jets.
		The vertical line indicates the CJV cut.
	}
\end{figure}

The effect of the CJV is illustrated in Fig.~\ref{fig:cenjetveto} where we show
the distribution of  the $p_T$ of the 3\textsuperscript{rd} light jet,
$p_{T_{j_3}}$, for the SM signal and the total background. Although the latter
has a harder spectrum  than the signal, imposing too stringent a veto is not
advantageous. This is because  the $b\bar{b}b\bar{b}$ final state leads to a 
non-negligible amount of hadronic activity in the central region, for instance
due to gluon radiation from the $b$ quarks and to $b$-jet misidentification.
Based on these results, in our analysis, we impose a CJV with the threshold value
reported in Table~\ref{tab:sel-cuts} and shown in the plots by the dot-dashed
line.

\subsection{Higgs reconstruction}
\label{sec:jet-sub}

The next step in our analysis is the reconstruction of the Higgs boson
candidates. This is done separately for each of the three event categories. In
the resolved category, starting with the six hardest $b$-jets in the event,
\footnote{Note that we start with the 6 hardest $b$-jets since
gluon splitting generates additional $b$-jets. Thus, there is a non zero probability
of missing signal $b$-jets (i.e. from Higgs decays) if we only restrict ourselves to the hardest 4.}
we reconstruct the first Higgs boson candidate $h_1$ by identifying it with the
pair of $b$-jets whose invariant mass is closest to the Higgs mass,
$m_h=125\,$GeV. Out of the remaining $b$-jet pairs, the one with an invariant
mass closest to $m_{h_1}$ is then assigned to be the second Higgs boson
candidate, $h_2$. In the case of the intermediate and boosted categories, each
of the mass-drop tagged jets is identified with a Higgs candidate. The second
Higgs candidate in the intermediate category is then formed by considering the
five hardest $b$-jets in the event and selecting the pair whose mass is closest
to $m_h$.

\begin{figure}[t]
	\centering
	\begin{minipage}{0.49\textwidth}\centering
		\hspace{1cm}LHC 14 [TeV]
		\includegraphics[width=\textwidth]{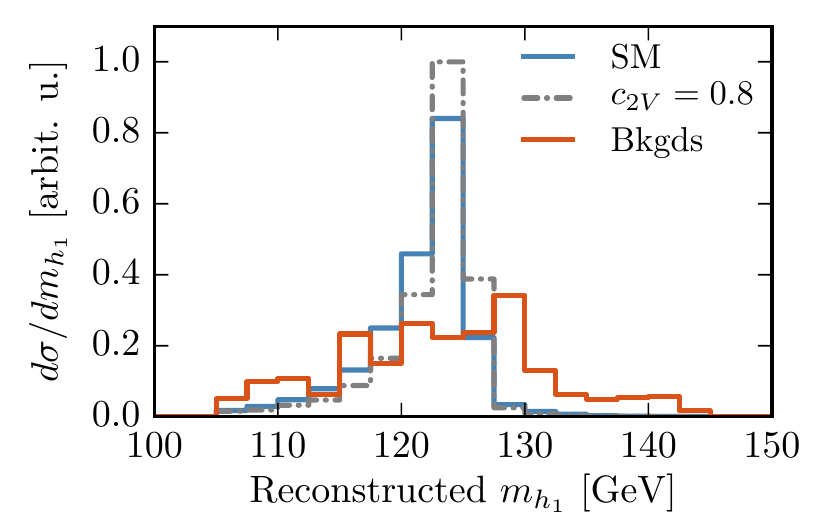}
		\includegraphics[width=\textwidth]{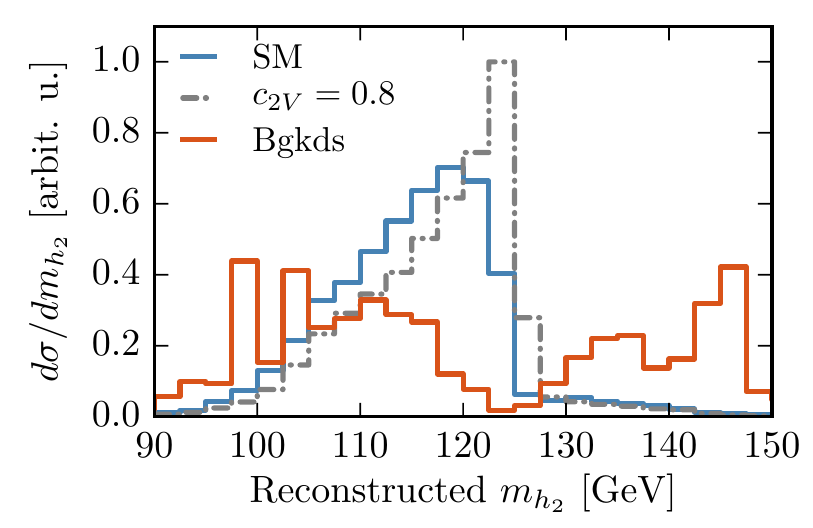}
	\end{minipage}
	\begin{minipage}{0.49\textwidth}\centering
		\hspace{1cm}FCC 100 [TeV]
		\includegraphics[width=\textwidth]{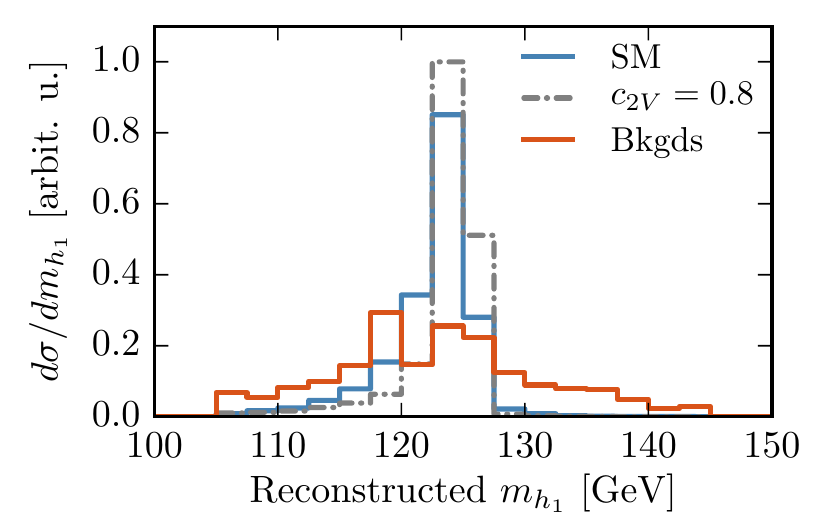}
		\includegraphics[width=\textwidth]{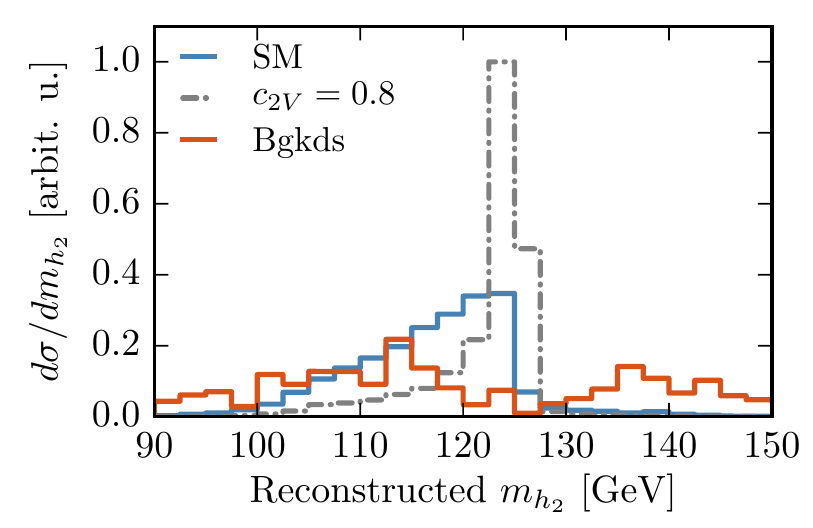}
	\end{minipage}
	\caption{\small \label{fig:higgsmass} 
	  Invariant mass distribution of the leading ($m_{h_1}$)
          and subleading ($m_{h_2}$) Higgs candidates 
		for signal (SM and $c_{2V}=0.8$) and background events
		at 14 TeV (left) and  100 TeV (right).
	}
\end{figure}

The invariant mass distributions of the Higgs candidates for the signal (SM and
$c_{2V}=0.8$) and the total background are shown in Fig.~\ref{fig:higgsmass}.
The peak around $m_h = 125\,$GeV is clearly visible for signal events,
especially in the case of $h_1$. The smearing of the signal distribution of the
second Higgs candidate $h_2$ arises from out-of-cone radiation effects which
reduce the reconstructed mass. It is largest in the SM, while it is reduced in
the $c_{2V}=0.8$ scenario and in particular at $100\,$TeV, due to the larger boost
of the Higgs bosons. The small peak in the background distributions for $h_1$ is
artificially sculpted by the analysis selection cuts. The fact that the
efficiency for the reconstruction of the Higgs bosons is similar in the SM and
for the $c_{2V}=0.8$ benchmark is another validation of the scale-invariant
tagging, since while in the SM  most of the events lead to resolved topologies,
the $c_{2V}=0.8$ scenario is dominated by the boosted category (see
Fig.~\ref{fig:mhh-cats} below).

After reconstructing the Higgs candidates, we require that their invariant masses, $m_{h_1}$ and
$m_{h_2}$, are reasonably close to the nominal mass.
%
In the resolved category, these conditions are:
\begin{align}
\label{eq:higgsmasswindow}
|m_{h_1}- 125\,\text{GeV}|~&\leq~20~{\rm GeV} \,,\\[0.2cm]
\label{eq:mh2cut}
|m_{h_2}-m_{h_1}|~&\leq~20~{\rm GeV} \, .
\end{align}
The mass window in Eq.~(\ref{eq:mh2cut}) is centered around  the mass of the
first candidate rather than the nominal Higgs mass in order to make the cut
robust against UE and PU effects. For the intermediate and boosted categories,
the invariant mass of all Higgs candidates is required to satisfy
Eq.~(\ref{eq:higgsmasswindow}). The cuts of
Eqs.~(\ref{eq:higgsmasswindow})-(\ref{eq:mh2cut})  are especially effective in
suppressing the QCD backgrounds which have almost featureless $m_{h_i}$
distributions in the Higgs mass regions. Finally, we impose an additional cut on
the invariant mass of the di-Higgs system:

\begin{equation}
\begin{split}
\text{LHC 14 TeV}&:\quad m_{hh}>500\phantom{0}\quad\text{GeV}\,,\\[0.2cm]
\text{FCC 100 TeV}&:\quad m_{hh}>1000\quad\text{GeV}.
\end{split}
\label{eq:mhhcut}
\end{equation}
This condition greatly reduces the background rates while leaving the
interesting kinematic region at large $m_{hh}$ -- where deviations from the SM signal
mostly appear -- unaffected.

\begin{figure}[tp]
	\centering
	\begin{minipage}{0.49\textwidth}\centering
		\hspace{1cm}LHC $14\,$TeV
		\includegraphics[width=\textwidth]{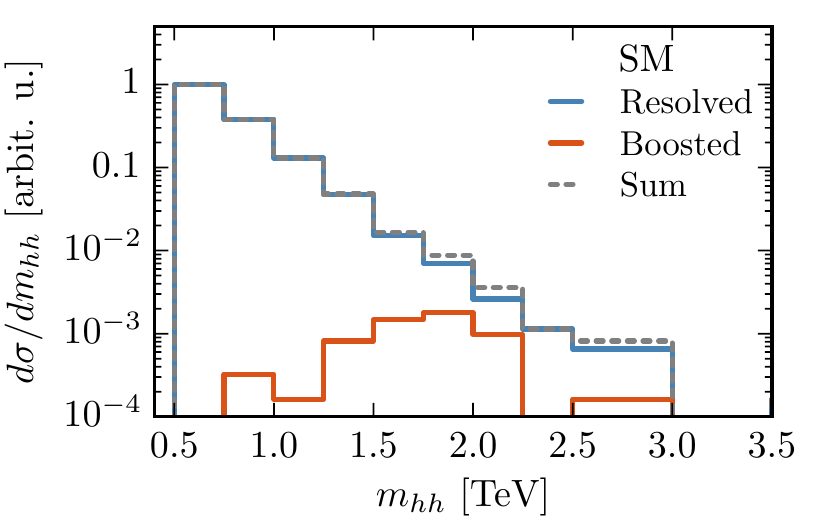}
		\includegraphics[width=\textwidth]{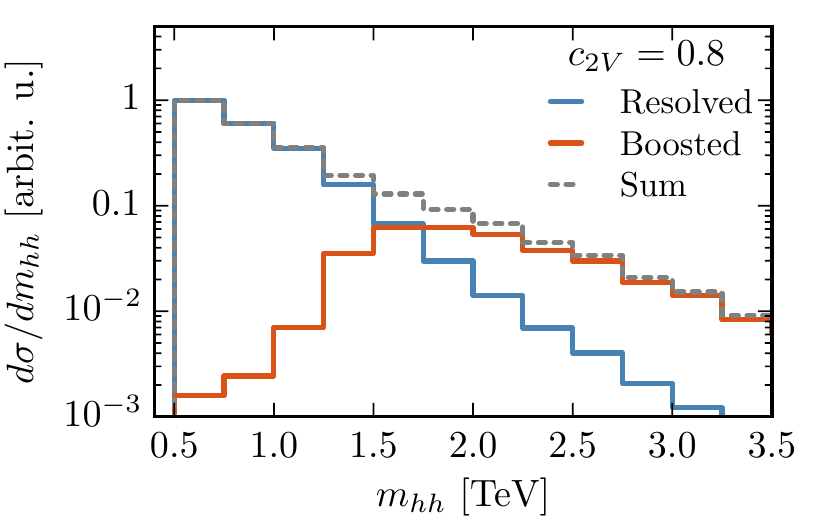}
		\includegraphics[width=\textwidth]{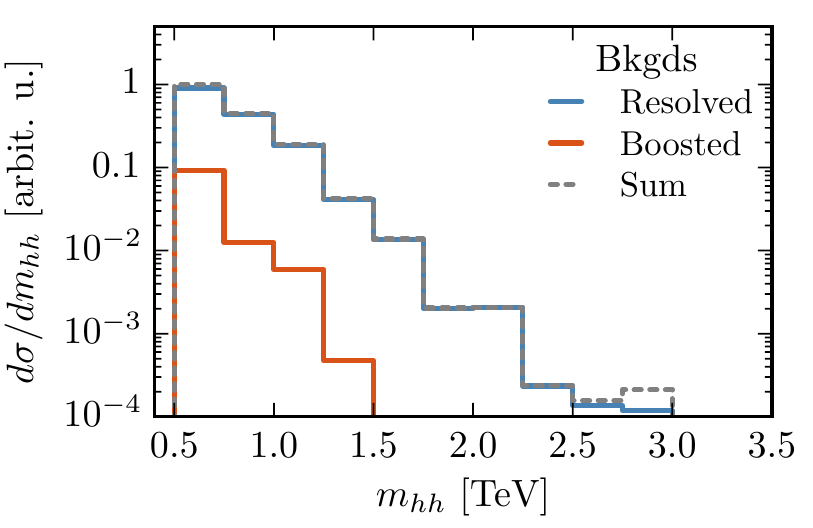}
	\end{minipage}
	\begin{minipage}{0.49\textwidth}\centering
		\hspace{1cm}FCC $100\,$TeV
		\includegraphics[width=\textwidth]{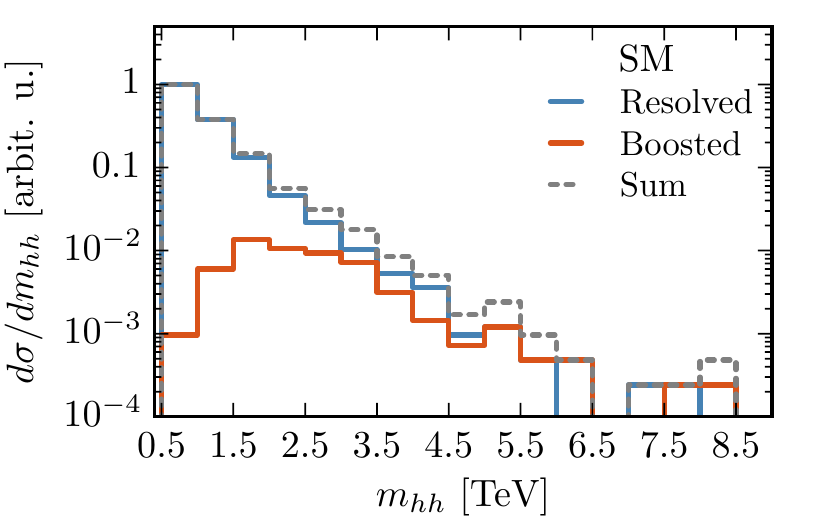}
		\includegraphics[width=\textwidth]{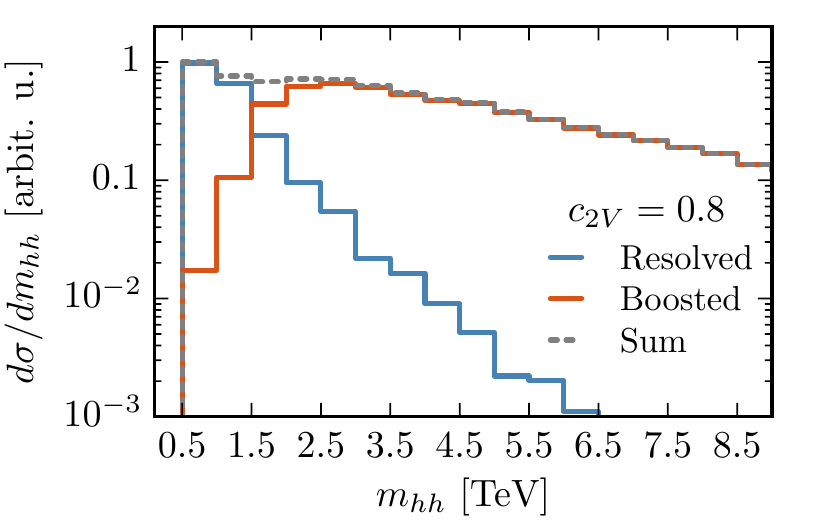}
		\includegraphics[width=\textwidth]{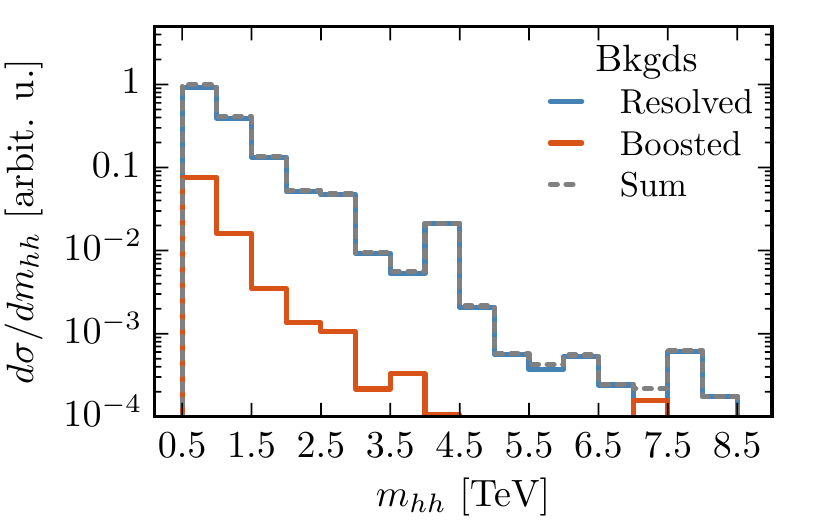}
	\end{minipage}
	\caption{\small \label{fig:mhh-cats} 
		Invariant mass distribution of the di-Higgs system
		at $14\,$TeV (left) and $100\,$TeV (right) after all analysis cuts,
                for the signal (SM and $c_{2V}=0.8$)
                and the total background.
              	We show the contribution
		from resolved and boosted events as well as the sum of the
		three categories.
	}
\end{figure}

Figure~\ref{fig:mhh-cats} shows the $m_{hh}$ distribution, after all the cuts listed in
Tables~\ref{tab:sel-cuts}~and Eqs.~(\ref{eq:higgsmasswindow})--(\ref{eq:mhhcut}),
for signal (SM and $c_{2V}=0.8$) and the total background at $14\,$TeV and
$100\,$TeV.
In each case, we show both the sum of the three event categories and the
individual contributions from  resolved and boosted events. The intermediate
category, which contributes very little in all cases, is not shown. This
comparison helps illustrate the relative weight of the boosted and resolved
categories. For signal events in the SM, the vast majority are classified in the
resolved category as  expected since in this case the boost of the di-Higgs
system is small except at 100 TeV and for large $m_{hh}$ values. On the
other hand, in the case of $c_{2V}=0.8$, the energy growth of the partonic cross
section induces a much harder $m_{hh}$ spectrum. This implies that, already at 14 TeV,
a substantial fraction of events falls in the boosted category which becomes
the dominant one at 100 TeV.
For $c_{2V}=0.8$, the crossover between the resolved and
boosted categories takes place
at $m_{hh}\simeq 1.5$ TeV for both colliders,
although this specific value depends on the choice of the jet
radius $R$~\cite{Gouzevitch:2013qca}.
Unsurprisingly, background events are always dominated by the resolved topology.

\subsection{Signal and background event rates}
\label{sec:rates}

Now that we have presented our analysis strategy, we can turn to discuss the
actual impact on the  cross sections and event rates of the various steps of the
cut flow. In Table~\ref{tab:xsecs} we report the cross sections at $14\,$TeV and
$100\,$TeV after acceptance, VBF, Higgs reconstruction, and $m_{hh}$ cuts of
Table~\ref{tab:sel-cuts} and Eqs.~(\ref{eq:higgsmasswindow})--(\ref{eq:mhhcut}),
respectively, for both the signal (SM and $c_{2V}=0.8$) and for the total
background.

\begin{table}[h]\centering
	\small
	\renewcommand{\arraystretch}{1.5}
	\begin{tabular}{llcccc}
	  \toprule[0.1em]
          & & \multicolumn{4}{c}{Cross-sections (fb) } \\
	&  &  Acceptance & VBF  & Higgs reco. & $m_{hh}$ cut  \\\midrule 
		\multirow{3}{*}{	{\large $14\,$TeV}} 
		& Signal SM &  0.011 & 0.0061 & 0.0039 & 0.0020 \\
		& Signal $c_{2V}=0.8$ & 0.035 & 0.020 & 0.017 & 0.011 \\
		 & Bkgd (total)   & $1.3\xtento{5}$ & $4.9\xtento{3}$ & 569 & 47   \\
		\midrule[0.04em]
		\multirow{3}{*}{	{\large $100\,$TeV}} 
		& Signal SM &  0.22 & 0.15 & 0.11 & 0.033 \\
		& Signal $c_{2V}=0.8$ & 3.4 & 2.7 & 1.9 & 1.6 \\
		 & Bkgd (total)   & $1.9\times 10^{6}$ & $1.9\times 10^{5}$ & $9.5\times 10^{3}$  &  212  \\
                \bottomrule[0.1em]
         \end{tabular} 
        \caption{\small Cross sections, in fb, at $14\,$TeV (upper table)
          and $100\,$TeV (lower table) after the
successive application of the acceptance and VBF cuts (Table~\ref{tab:sel-cuts}) and of the Higgs reconstruction cuts
(Eqs.~(\ref{eq:higgsmasswindow})-(\ref{eq:mhhcut})), for signal events (SM and  $\cvv=0.8$) and for
the total background. \label{tab:xsecs}
}
\end{table}

At $14\,$TeV, we find that the VBF di-Higgs signal in the SM is rather small
already after the basic acceptance cuts. On the other hand, the signal event
yield is substantially increased for $c_{2V}\ne 1$ as illustrated by the
benchmark value of $\cvv = 0.8$ leading to more than a factor~3\,(5) enhancement
compared to the SM after the acceptance (all analysis) cuts. The fact that this
cross-section enhancement for the $c_{2V} = 0.8$ scenario is more marked at the
end of the analysis is not a coincidence: our selection cuts have been designed
so as to improve the sensitivity to $c_{2V}$ by increasing the signal
significance in the large-$m_{hh}$ region.

From Table~\ref{tab:xsecs} we also find that a similar qualitative picture holds
at $100\,$TeV with the important difference that, in this case, the event rate
is already substantial in the SM which yields $\simeq 2000$ events after the
acceptance cuts with $\mathcal{L}=10\,$ab$^{-1}$. The cross section enhancement
at $100\,$TeV as compared to $14\,$TeV is driven by the larger centre-of-mass
energy and leads to a signal rate greater by a factor 20\,(17) after the
acceptance (all analysis) cuts in the SM, and by a factor $\simeq 100\,(150)$ in
the $c_{2V}=0.8$ scenario. At $100\,$TeV, the ratio of signal cross sections in
the $c_{2V}=0.8$ and SM scenarios is $\sim 15\,(50)$ after acceptance (all
analysis) cuts. Note however that at both $14\,$TeV and $100\,$TeV, even after
all analysis cuts the background is still much larger than the signal (either SM
or $c_{2V}=0.8$) at the level of inclusive rates. It is only by exploiting  the
large-$m_{hh}$ region that the former can be made small enough to achieve high
signal significances.

Table~\ref{tab:xsecs} is also useful to assess the relative impact on the signal and
the total background of each of the cuts imposed.
In the case of the VBF cuts, we find that the background is drastically reduced,
by more than one order of magnitude, at the cost of a moderate decrease of the signal
cross sections.
The Higgs mass window requirement is also instrumental to further suppress the backgrounds, especially
the QCD multijets which are featureless in $m_h$, while leaving the signal
mostly unaffected.
A final reduction of the background, by around another order of magnitude,
is achieved through the $m_{hh}$ cut.
The relative impact of each cut on signal events
is similar in the SM and for $c_{2V}=0.8$.

\begin{figure}[t]
	\centering
		\includegraphics[width=0.49\textwidth]{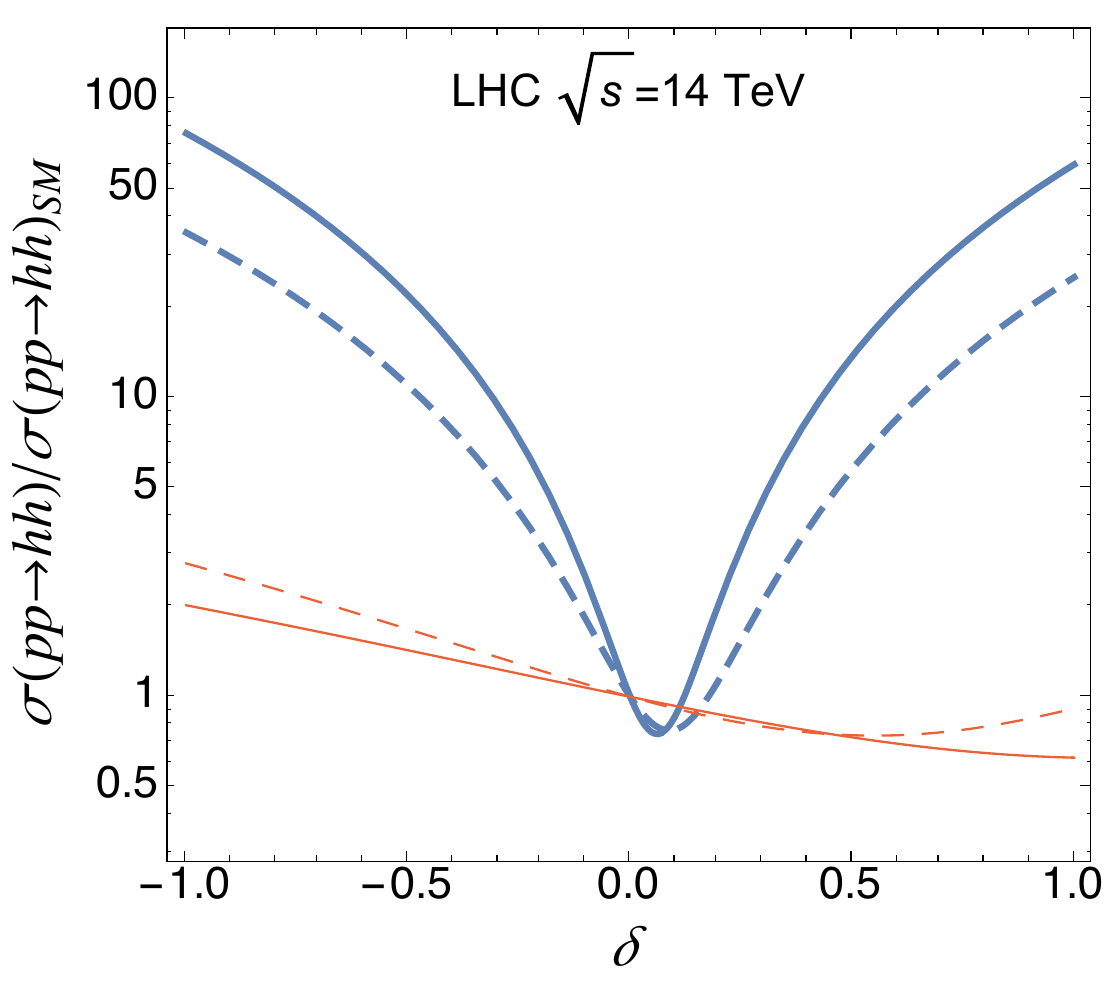}
		\includegraphics[width=0.49\textwidth]{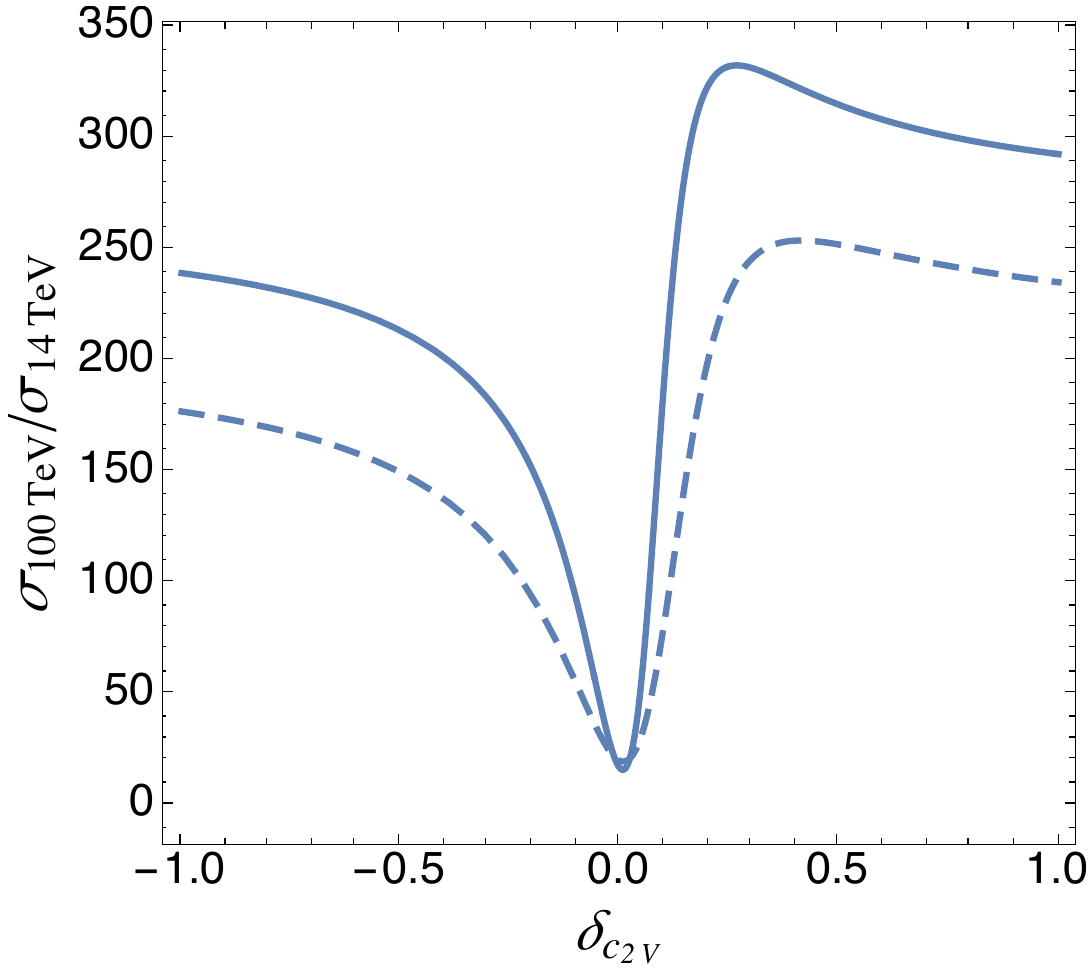}
                \vspace{-0.3cm}
                \caption{\small Left panel: VBF di-Higgs cross-section, in units of the
                  SM value, as a function of $\dcvv$ (thick blue)
                  and $\delta c_{3}$ (thin red), after
                  acceptance cuts (solid)
                  and all analysis cuts (dashed).
                  Right panel: ratio of VBF di-Higgs cross section between $100$ TeV and
                  14 TeV as a function of $\dcvv$.
                  \label{fig:xsec_vs_delta}
                }
\end{figure}

Figure~\ref{fig:xsec_vs_delta} graphically illustrates the dependence of the di-Higgs production cross section
on the couplings $c_{2V}$ and $c_{3}$. 
%
%
The left panel shows the cross section in SM units as a function of
$\dcvv=c_{2V}-1$ and $\delta_{c_{3}}=c_{3}-1$  after applying the acceptance
cuts of Table~\ref{tab:sel-cuts} (dashed curves) and after all the analysis cuts
(solid curves). The sensitivity on $c_{2V}$ is particularly striking, for
example the cross section for $|\dcvv|\simeq 1$ is enhanced by a factor $\sim
50$ compared to its SM value after all cuts. This sensitivity is the key
ingredient for measuring $\cvv$ with good precision, even though the SM cross
section itself cannot be extracted with comparable accuracy. In particular, as
we will show in Sec.~\ref{sec:results}, the sensitivity to $\dcvv$ derives
mainly from the tail of the $m_{hh}$ distribution. This observation elucidates
the enhancement (suppression) in the sensitivity to $\dcvv\,(\dccc)$ in
Fig.~\ref{fig:xsec_vs_delta} after the application of all the cuts which, for
instance, remove the threshold region up to $m_{hh}=500\,(1000)$ GeV at the
LHC(FCC).

The right panel of Fig.~\ref{fig:xsec_vs_delta} shows, instead, the ratio
between the VBF di-Higgs cross sections at $\sqrt{s}=100\,$TeV and $14\,$TeV.
Given the larger centre-of-mass energy of the FCC, it is expected that this
ratio grows rapidly for $\dcvv \ne 0$ and, indeed, it can reach values as high
as 300 for $\dcvv\simeq 1$. As will be demonstrated in Sec.~\ref{sec:results},
this effect allows for much more precise measurements of $c_{2V}$ at the FCC,
with uncertainties reduced by a factor 20 as compared to the HL-LHC. The results
of Fig.~\ref{fig:xsec_vs_delta} are of course consistent with the findings of
Table~\ref{tab:xsecs}.
                  
From Fig.~\ref{fig:xsec_vs_delta}, we also observe that the sensitivity of the
signal on the Higgs trilinear coupling $c_3$ is relatively weak even for large
variations and it is reduced by our analysis strategy. As already mentioned,
this latter feature is expected because the sensitivity to~$c_{3}$ comes from
events near the di-Higgs threshold, $m_{hh}\simeq 2m_{h}$, which are removed by
our cuts due to the overwhelming backgrounds in that region. This weak
dependence of the VBF di-Higgs cross-section on $c_3$, together with the large
event rates for background after all the analysis cuts (see
Table~\ref{tab:xsecs}), suggest that the VBF process is not suitable to extract
the Higgs self-coupling.

Let us now turn to discuss the background processes. As mentioned above and
discussed in~\ref{sec:mc}, there are two types of processes that
contribute to the final-state signature under consideration. The first type are
QCD processes and in particular multijet and top-quark pair production in
association with additional hard radiation. The second is Higgs pair production
in the gluon-gluon fusion channel in association with additional jets, where the
latter can mimic the VBF topology, as in single-Higgs production.

In the case of QCD multijet processes,
it is important to account for the effects of both the $4b$ and the $2b2j$ backgrounds (where
we label each process by its matrix-element level content; as explained
in~\ref{sec:mc}, additional jets are generated by the parton shower).
The latter process can lead to events being classified as signal when
light jets are misidentified as $b$-jets or when a gluon splits into a $b\bar b$ pair during the parton shower.
Even with a small light jet mistag rate of $\mcO(1\%)$, it can have
a contribution to the total background comparable to or bigger than the $4b$ process.
Details on the generation of the QCD backgrounds and on the associated
validation tests  are presented in~\ref{sec:mc} and~\ref{sec:qcdmultijet}.

\begin{table}[t]\centering
	\small
	\renewcommand{\arraystretch}{1.5}
	\begin{tabular}{lccccc}
		\toprule[0.1em]
	 &   &  Acceptance & VBF  & Higgs reco. & $m_{hh}$ cut  \\\midrule 
\multirow{5}{*}{\large LHC 14 TeV}
& $4b$ &  $ 1.18\xtento 4 $ & 613 & 54 & 4.45 \\
& $2b2j$  & $ 1.14\xtento 5 $ & $ 4.31\xtento 3 $ & 514 & 42.6 \\
& $t\bar{t}jj$  & 150 & 4.75 & 0.732 & 0.0706 \\
& $gg\to hh$  & 0.98 & 0.0388 & 0.0223 & 0.00857 \\
& Total   & $1.3\xtento{5}$ & $4.9\xtento{3}$ & 569 & 47   \\
\midrule
\multirow{5}{*}{\large FCC 100 TeV}
& $4b$ &  $ 3.93\xtento 5 $ & $ 4.59\xtento 4 $ & $ 2.61\xtento 3 $ & 106 \\
& $2b2j$ &  $ 1.52\xtento 6 $ & $ 1.46\xtento 5 $ & $ 6.88\xtento 3 $ & 104 \\
& $t\bar{t}jj$  & $ 9.76\xtento 3 $ & 832 & 55 & 1.47 \\
& $gg\to hh$  & 24.8 & 2.48 & 1.31 & 0.0892 \\
& Total &   $1.9\times 10^{6}$ & $1.9\times 10^{5}$ & $9.5\times 10^{3}$  &  212  \\
\bottomrule[0.1em]
\end{tabular}
\caption{\small Same as Table~\ref{tab:xsecs}, now listing separately each background process. }
\label{tab:xsecsBack}
\end{table}

Concerning gluon-fusion Higgs pair production  in association with additional
hard jets,  similarly to single-Higgs VBF production there will be certain
configurations that mimic the VBF topology as emphasized, for example, in
Ref.~\cite{Dolan:2015zja}. In contrast to the VBF channel, however, the Higgs
pair production in gluon-fusion does not exhibit any enhancement in
the tail of the $m_{hh}$ distribution. This substantially
reduces its contamination to the region with the highest sensitivity to $\cvv$
in our analysis. Note also that a harder $m_{hh}$ distribution could be
generated by higher-order EFT operators, for instance those leading to a contact
interaction of the form $gghh$, as in Ref.~\cite{Azatov:2015oxa}. The
investigation of this scenario is however left for future work.

In Table~\ref{tab:xsecsBack}, following the structure of Table~\ref{tab:xsecs},
we give the cross sections at $14\,$TeV and $100\,$TeV for the individual
background processes (and their sum) after the acceptance, VBF, Higgs
reconstruction, and $m_{hh}$ cuts. We find that in all steps in the cut-flow the
dominant background component is QCD multijet production, both at  $14\,$TeV and
at $100\,$TeV. After all analysis cuts, the $2b2j$ component is a factor 10
larger than $4b$ at 14 TeV while they are of similar size at 100 TeV.

Other backgrounds, including gluon-fusion di-Higgs production, are much smaller
than QCD multijets. Note however that the former is actually larger than the VBF
signal for SM couplings with a cross-section at 14 TeV of $0.98 \, (0.009)\,$fb
after acceptance (all) cuts, compared to $0.11 \, (0.002)\,$fb for the VBF case.
On the other hand, this fact does not affect the measurement $c_{2V}$ since, as
we show next, the sensitivity comes from the large $m_{hh}$ tail where the
gluon-fusion component is heavily suppressed.

\begin{figure}[t]
	\centering
	\begin{minipage}{0.49\textwidth}\centering
		\hspace{1cm}LHC $14\,$TeV
		\includegraphics[width=\textwidth]{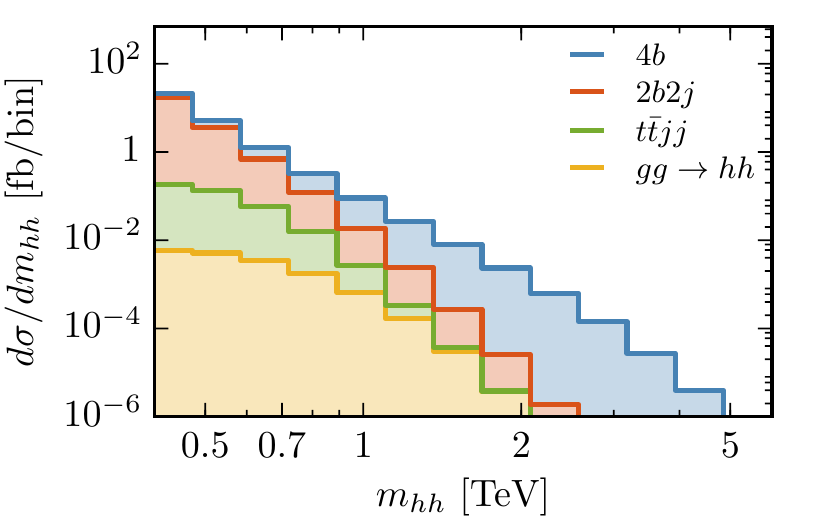}
	\end{minipage}
	\begin{minipage}{0.49\textwidth}\centering
		\hspace{1cm}FCC $100\,$TeV
		\includegraphics[width=\textwidth]{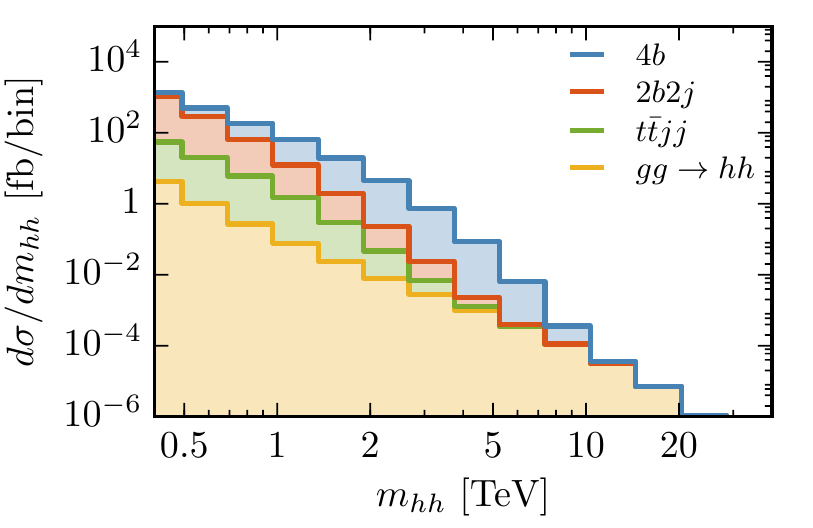}
	\end{minipage}
	\caption{\small Decomposition of the total
          background into individual processes as a function the di-Higgs
          invariant mass
          after all analysis cuts have been imposed,
          except for the $m_{hh}$ cut.
	}
\label{fig:bkg-comp}
\end{figure}

The decomposition of the total background in terms of individual processes as a
function of $m_{hh}$ is shown in Fig.~\ref{fig:bkg-comp} where the components are
stacked on top of each other so that the content of each bin matches the total
background cross section. In the $14\,$TeV case, the $4b$ background dominates
for large $m_{hh}$ while the $2b2j$ one is instead the most important for small
$m_{hh}$. The $100\,$TeV case is similar with one exception, namely that the
gluon fusion di-Higgs background becomes the dominant one for very high
invariant masses, $m_{hh}\gsim 10\,$TeV. Such extreme region is however
phenomenologically irrelevant due to the very small rates of both signal and
background even at a 100 TeV collider.

%% file: sec-results.tex
\section{Results}
\label{sec:results}

The last column of Table~\ref{tab:xsecs} indicates
the cross sections for the signal 
and total background after imposing all analysis cuts.
We observe that the background, dominated by QCD multijets,
still has a much larger cross section than the signal,
both in the SM and in the $c_{2V}=0.8$ benchmark scenario.
As anticipated, the additional handle which we can now
exploit to increase the signal
significance is  the 
different behaviour of the
$m_{hh}$ distribution for the signal and the background,
in particular
when $c_{2V}$ departs from its SM value. 
The latter has a sharp fall-off
at large $m_{hh}$ values while, instead, the signal exhibits
a much harder spectrum for $c_{2V} \not = 1$.
This  cross-section growth implies that,
for $|\dcvv|$ sufficiently large, there will be
a crossover value of $m_{hh}$ where the signal
overcomes the background. 

\begin{figure}[tp]
	\centering
	\begin{minipage}{0.49\textwidth}\centering
		\hspace{1cm}LHC $14\,$TeV
		\includegraphics[width=\textwidth]{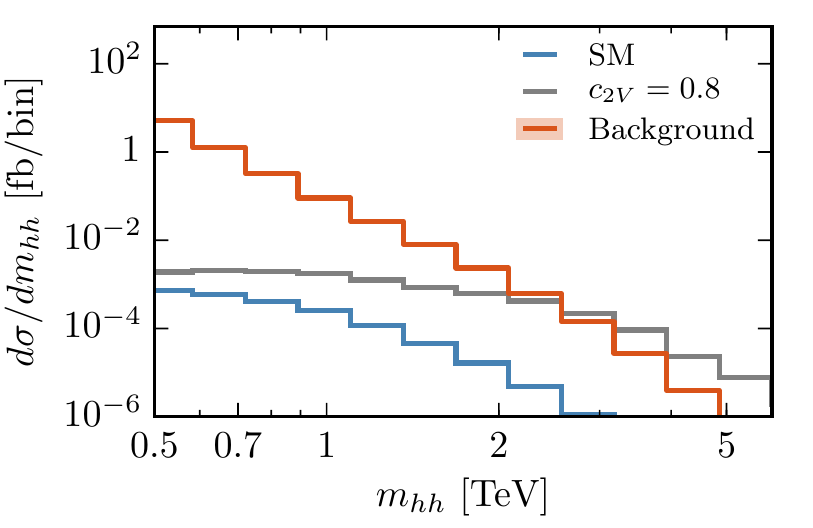}
	\end{minipage}
	\begin{minipage}{0.49\textwidth}\centering
		\hspace{1cm}FCC $100\,$TeV
		\includegraphics[width=\textwidth]{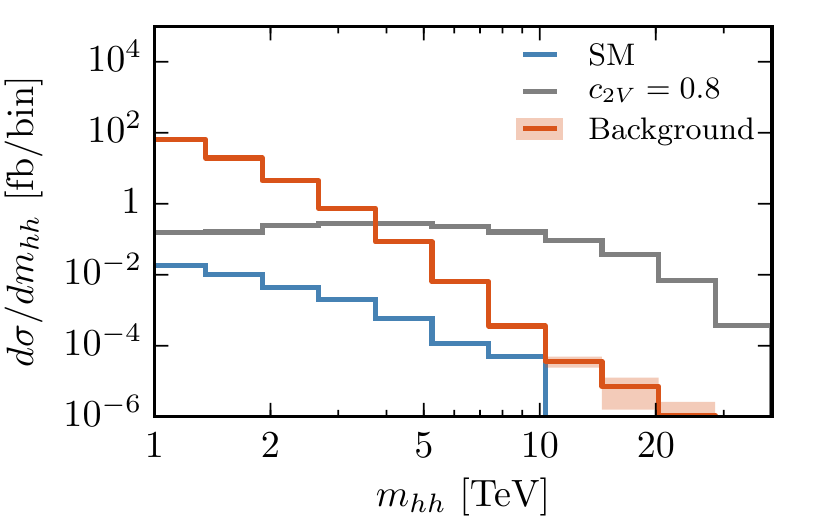}
	\end{minipage}
	\caption{\small \label{fig:mhh-abs} 
	  The di-Higgs $m_{hh}$ distribution at 14 TeV (left)
          and 100 TeV (right)  after all analysis cuts
          showing the results for the signal (SM and $c_{2V}=0.8$) and for the total background.
      }
\end{figure}

This behaviour
is illustrated in
Fig.~\ref{fig:mhh-abs} where we show
the invariant mass distribution of the Higgs pairs  after all analysis cuts, at $14\,$TeV 
and $100\,$TeV, 
for  the signal (SM and $c_{2V}=0.8$)
and the total background.
In the case of the benchmark scenario with $c_{2V}=0.8$,
the 
crossover between signal and background
is located  at $m_{hh}\simeq 2\,$TeV ($4\,$TeV) at
$14\,$TeV ($100\,$TeV).
We also observe that,
for invariant masses $m_{hh}$ above this crossover,
the ratio between
the signal and the backgrounds keeps increasing steeply.

With the final results of our analysis in hand,  we can now estimate the expected
sensitivity to deviations in the $hhVV$ coupling, parametrized as
$\dcvv=\cvv-1$, by exploiting the information
contained in the full  $m_{hh}$ differential distribution (as opposed to using
only the total
number of events satisfying all cuts from Table~\ref{tab:xsecs}).
To achieve this,
we first bin our results in $m_{hh}$ and then
follow a Bayesian approach~\cite{D'Agostini:2003nk}
to construct a posterior probability density function.
We include two nuisance parameters, $\theta_B$ and $\theta_S$,
to account for the uncertainty associated with the
background and 
signal event rate, respectively.
The parameter $\theta_S$ encodes the 
theoretical uncertainties on the di-Higgs cross section and
the branching fraction ${\rm BR}(h\to b\bar{b})$.
We conservatively assume a $10\%$ uncertainty uncorrelated in each $m_{hh}$ bin.


Concerning $\theta_B$,
we expect that an  actual experimental analysis of di-Higgs production via VBF
would estimate the overall normalization of the 
different background components by means of data-driven techniques.
We assume a 15\% uncertainty
arising from the measurement and subsequent extrapolation
of the dominant QCD multijet background, see for example
a recent ATLAS measurement of dijet $b\bar{b}$
cross-sections~\cite{Aaboud:2016jed}.
The background nuisance parameter,
$\theta_B$, is conservatively also
assumed to be uncorrelated
among $m_{hh}$ bins.
In addition, while we already
rescale the background cross sections to match existing
NLO and NNLO results (see~\ref{sec:mc}),
there still remains a sizable uncertainty in
their overall normalization from missing higher orders,
in particular for the QCD multijet components.
For this reason, below, we explore the robustness of our results
upon an overall rescaling of all the background
cross sections by a fixed factor.

The posterior probability function constructed in this way reads:
\begin{equation}
  \label{eq:postpdf1}
  P(\dcvv|\{N^i_{\rm obs}\})=\int \!  \prod_{i\in\{\text{bins}\}}\!
  d\theta_S^i \, d\theta_B^i\;
  L\!\left(N^i(\theta_B^i,\theta_S^i)|N_\text{obs}^i\right) 
		\, e^{-(\theta_S^{i})^2/2}\, e^{-(\theta_B^{i})^2/2} \, \pi(\cvv)\, ,
\end{equation}
with $N^i(\theta_B^i,\theta_S^i)$ and $N^i_\text{obs}$ denoting respectively the number of
predicted (for a generic value of $\cvv$) and observed 
(assuming SM couplings) events for a given
integrated luminosity ${\cal L}$ in the $i$-th bin
of the di-Higgs invariant mass distribution $m_{hh}$, given by
\footnote{\label{fn:bin-def}In our analysis, we use 15 bins starting at 250 GeV up to 6(30)TeV for the LHC(FCC) that are uniformly spaced on a log scale. In addition, we define an overflow bin up to the relevant centre of mass energy.}:
\begin{equation} \label{eq:postpdf}
\begin{split}
	N^i(\theta_B,\theta_S)&=\left[ \sigma_\text{sig}^i(\cvv)\left(1+\theta_S^i\,\delta_S\right)+
	  \sigma_\text{bkg}^i\left(1+\theta_B^i\,\delta_B\right) \right] \times {\cal L}\, ,
        \\[0.2cm]
	N_\text{obs}^i&=\left[ \sigma_\text{sig}^i(\cvv=1)+\sigma_\text{bkg}^i \right] \times {\cal L}\, .
\end{split}
\end{equation}
In Eq.~(\ref{eq:postpdf}),
$\sigma_\text{sig}^i(\cvv)$ and $\sigma_\text{bkg}^i$ indicate
the signal (for a given
value of $\cvv$) and total
background cross sections, respectively, for the $i$-th bin of the $m_{hh}$ distribution.
The functional form of $\sigma_\text{sig}^i(\cvv)$ is given by Eq.~(\ref{eq:xsecgeneral}) and the value of the coefficients in bin $i$ are given in~\ref{sec:fit-coeffs}.
We denote by $\pi(\cvv)$ the prior probability distribution
of the $\cvv$ coupling.

\begin{figure}[t]
	\centering
	\begin{minipage}{0.49\textwidth}\centering
		\includegraphics[width=\textwidth]{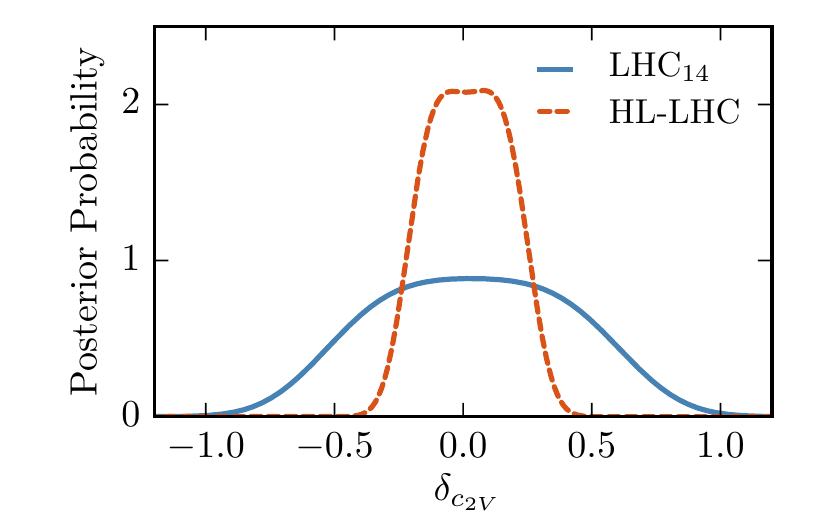}
	\end{minipage}
	\begin{minipage}{0.49\textwidth}\centering
	\includegraphics[width=\textwidth]{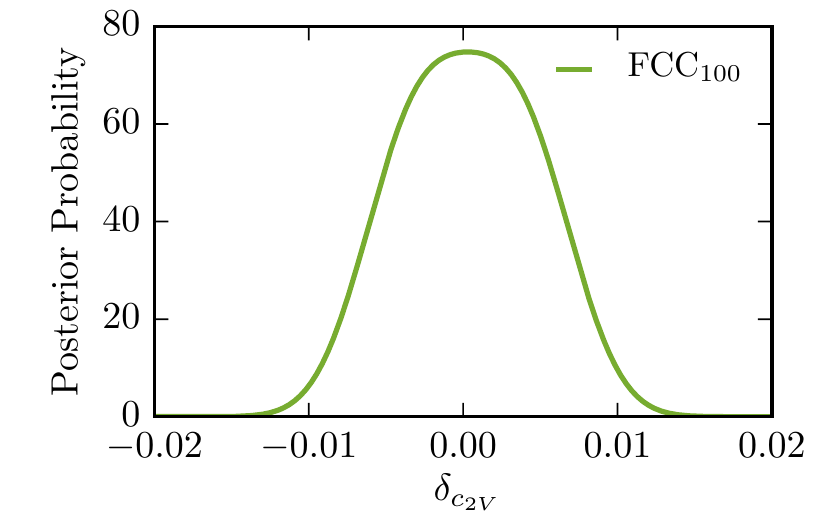}
	\end{minipage}
	\caption{\small
	  Posterior probability densities for $\dcvv$ at the LHC
   for $\mathcal{L}=300$ fb$^{-1}$ (LHC$_{14}$) and
       $\mathcal{L}=3$ ab$^{-1}$  (HL-LHC)
   and for the FCC with $\mathcal{L}=10$ ab$^{-1}$.
 Note the different scales of the axes in the two panels. 
	}
	\label{fig:post-pdf} 
\end{figure}

As justified above, in the evaluation of
Eq.~(\ref{eq:postpdf1}) we set
$\delta_{B(S)} = 0.15\;(0.1)$ and assume that the two
nuisance parameters are normally distributed.
We have verified that assuming instead
a log normal distribution leads to similar results.
  In addition, we take a Poissonian likelihood $L(N^i|N^i_\text{obs})$ in
  each bin and assume the prior probability $\pi(c_{2V})$ to be uniform.
The resulting
posterior probabilities are shown in Fig.~\ref{fig:post-pdf}
for the LHC
   with $\mathcal{L}=300$ fb$^{-1}$ (LHC$_{14}$) and
       $\mathcal{L}=3$ ab$^{-1}$  (HL-LHC),
   and for the FCC with $\mathcal{L}=10$ ab$^{-1}$.
To produce this figure, as well as to determine the values reported in Tabs.~\ref{tab:resultsdcvv} and~\ref{eq:results_ss}, we included all bins with at least one event.

From Fig.~\ref{fig:post-pdf}, we can determine the expected precision  for a
measurement of $\dcvv$ at the LHC  and the FCC in the case of SM values of the
Higgs couplings.
The 68\% probability intervals for the determination of $\cvv$ at the LHC and
the FCC are listed in Table~\ref{tab:resultsdcvv}. This is the central result of
this work.
To assess its robustness with respect to our estimate of the background cross
sections, we also provide the same intervals in the case of an overall rescaling
of the total background by a factor 3.
Furthermore, we can also assess the effect of varying $\cv$ on the bound on $\dcvv$ by
treating $\cv$ as a nuisance parameter and marginalizing over it.
The leading effect of varying $\cv$ comes from the $(\cvv-\cv^2)$ term at the amplitude level -- see Eq.~\ref{eq:growth} -- and can be included using the parametrization of Eq.~\ref{eq:xsecprime}. The neglected
dependence is sub-leading and arises from the interference of diagrams proportional to $\cv^2$ and $\cv\ccc$.
We take $\cv$ to be Gaussian distributed with a mean equal to 1 (i.e., its
SM value) and a width equal to 4.3\%, 3.3\%, and 2\% at the LHC Run II, HL-LHC, and FCC respectively.
In case of the LHC (both Run II and HL), the width of the Gaussian
corresponds to the projected sensitivity from the two parameter fit by ATLAS~\cite{ATL-PHYS-PUB-2014-016}.
The effect of marginalizing over $\cv$ is sub-leading in both LHC scenarios and weakens
the bound on $\dcvv$. 
We find that the results of Table~\ref{tab:resultsdcvv} change by 2\% for LHC$_{14}$ and 7\% for HL-LHC.
The effect
at the FCC is much larger causing the bound on $\dcvv$ to be $\mathcal{O}(0.04)$ rather than 0.01. This is not surprising and indicates that a joint likelihood would be required
at the FCC.
\begin{table}[t]
\begin{center}
  \begin{tabular}{r|@{\hskip 0.15in}c @{\hskip 0.2in}c}
  	\toprule[1pt]
  	&\multicolumn{2}{c}{68\% probability interval on $\dcvv$}\\[0.1cm]
    & $1\times\sigma_\text{bkg}$&
    $3\times\sigma_\text{bkg}$ \\[0.1cm]
\hline
 & & \\[-0.3cm]
LHC$_{14}$ & [$-0.37$,\,$0.45$] & [$-0.43$,\,0.48] \\[0.2cm]
HL-LHC & [$-0.15$,\,0.19] & [$-0.18$,\,0.20]\\[0.2cm]
FCC$_{100}$ & [$0$,\,0.01] & [$-0.01$,\,0.01]  \\
\bottomrule[1pt]
\end{tabular}
\end{center}
\vspace{-0.3cm}
\caption{\small Expected precision (at 68\% probability level) for
  the measurement of $\dcvv$ at the LHC  and the FCC, assuming
  SM values of the Higgs couplings.
  We show results both for the nominal background cross section
  $\sigma_\text{bkg}$, and for the case in which this value is rescaled  by a factor 3.
  \label{tab:resultsdcvv}
}
\end{table}

From Table~\ref{tab:resultsdcvv}, we find that the $c_{2V}$ coupling, for which
there are currently no direct experimental constraints, can already be measured at the
LHC with $300\,\text{fb}^{-1}$ with a reasonably good accuracy:
$_{-37\%}^{+45\%}$ with 68\% probability. This accuracy is only marginally
degraded if the background is increased by a factor 3. A better precision, of
the order of  $_{-15\%}^{+19\%}$, is expected at the HL-LHC with
$3\,\text{ab}^{-1}$. Also, this estimate is robust against an overall rescaling
of the background cross section. Finally, we find a very significant improvement
at the FCC with $10\,\text{ab}^{-1}$, where a measurement at the 1\% level could
be achieved providing an unprecedented test for our understanding of the Higgs
sector.

It is interesting to compare these results with the experimental precision
expected on the fiducial VBF di-Higgs production cross section after all
analysis cuts, expressed in terms of the signal strength parameter normalized to
the SM result, $\mu = \sigma/\sigma_\text{\sc sm}$. Table~\ref{eq:results_ss}
shows the 95\% upper limits on $\mu$ for the nominal background cross section and
after rescaling the latter by a factor 3. The comparison with
Table~\ref{tab:resultsdcvv} clearly shows that the high precision expected on
$c_{2V}$ can obtained despite the rather loose constraints that can be obtained
on the VBF di-Higgs cross section even at $100\,$TeV. As already discussed,
this behaviour follows from the strong dependence of the signal cross section on
$c_{2V}$, see Fig.~\ref{fig:xsec_vs_delta}.

\begin{table}[t]
  \begin{center}
      \begin{tabular}{r|@{\hskip 0.08in} cc}
      	\toprule[1pt]
	& \multicolumn{2}{c}{95\% probability upper limit on $\mu$}\\[0.1cm]
    & \hspace{0.22in} $1\times \sigma_\text{bkg}$ & $3\times\sigma_\text{bkg}$ \\[0.1cm]
\hline
 & & \\[-0.3cm]
LHC$_{14}$ & 109 & 210 \\[0.2cm]
HL-LHC & 49 & 108 \\[0.2cm]
FCC$_{100}$ & 12 & 23  \\
\bottomrule[1pt]
\end{tabular}
  \end{center}
\vspace{-0.3cm}
\caption{\small 95\% probability upper limits on
  the fiducial signal strength, $\mu = \sigma/\sigma_\text{\sc sm}$.
  \label{eq:results_ss}
}
\end{table}

The results of Tables~\ref{tab:resultsdcvv} and~\ref{eq:results_ss}
have been obtained by making full use of the information
contained on the di-Higgs invariant mass distribution $m_{hh}$.
However, the EFT expansion might break down at large enough values of
$m_{hh}$, corresponding to large partonic center-of-mass energies,
and some assessment on the validity of our procedure is thus required.
In particular, results can be consistently derived within the EFT
framework only if the new physics scale $\Lambda$ is smaller than the largest value of $m_{hh}$ included in the analysis.

As stressed in Ref.~\cite{Contino:2016jqw}, constraining $\Lambda$
requires making assumptions on the structure of the UV dynamics extending the SM.
For example, for the case where the new physics is characterized by a single 
coupling strength~$g_*$ and mass scale $\Lambda$~\cite{Giudice:2007fh}, one naively expects
\begin{equation} \label{eq:NDA}
\dcvv \approx g_*^2 v^2/\Lambda^2\, .
\end{equation}
Therefore, for maximally strongly-coupled UV completions (with $g_* \simeq 4\pi$)
it is possible to derive the following upper limit,
\begin{equation}
\label{eq:dcvvmax}
\dcvv^\text{max} \approx 16\pi^2 v^2/\Lambda^2 \, ,
\end{equation}
which makes explicit the connection between the value of $\dcvv$ and the new physics scale~$\Lambda$.
The validity of the EFT can thus be monitored
by introducing a restriction on the $m_{hh}$ bins
used in the construction of the posterior probability Eq.~(\ref{eq:postpdf1}),
so that $m_{hh} \leq  m_{hh}^\text{max}$, and
then determining how the sensitivity on $\dcvv$ varies as a function of 
$m_{hh}^\text{max}$~\cite{Contino:2016jqw}.

\begin{figure}[t]\centering
  \includegraphics[width=0.49\textwidth]{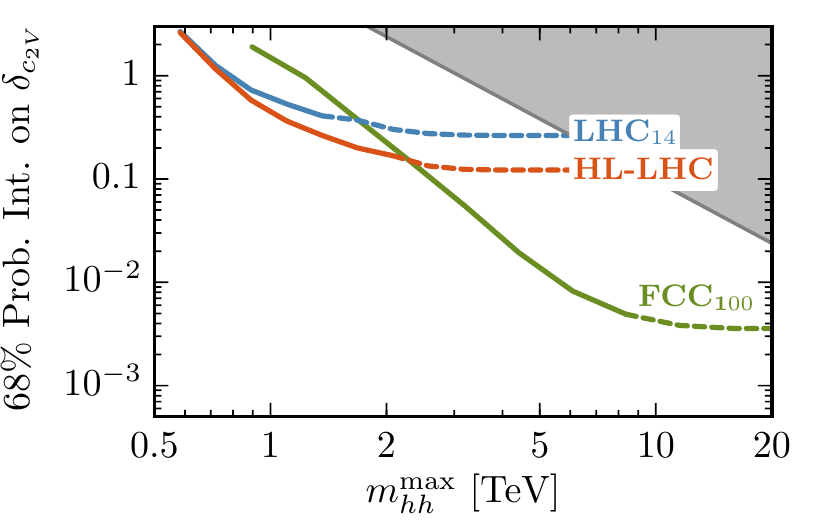}
  \includegraphics[width=0.49\textwidth]{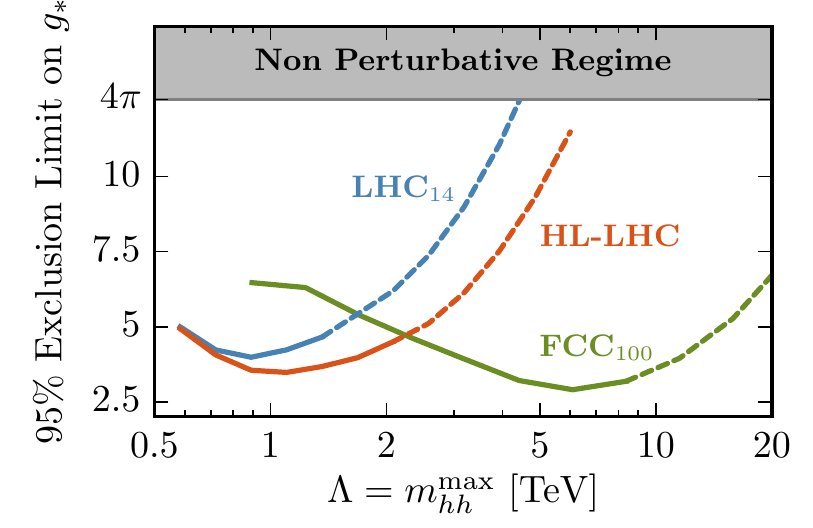}
  \caption{\small Left: the expected
    precision for a measurement of $\dcvv$ at the 68\% CL
          as a function of $m_{hh}^{\max}$ for
          the LHC and the FCC.
          The gray area indicates the region where $\dcvv > \dcvv^\text{max}$,
          obtained by setting 
          $\Lambda = m_{hh}^\text{max}$.
          Right: the 95\% CL exclusion limits in the $(\Lambda=m_{hh}, g_*)$ plane,
          assuming  Eq.~(\ref{eq:NDA}),
          where again the gray area corresponds to the
          non-perturbative regime, defined by $g_* \ge 4\pi$.
          The transition between solid and dashed curves occurs at the last bin with at least one event.
}
	\label{fig:c2v-sens}\label{fig:ms-gs}
\end{figure}

The precision on ${\dcvv}$, defined though the symmetrized 68\% probability interval,
is shown in Fig.~\ref{fig:c2v-sens}~as a function of  $m_{hh}^{\max}$ for the LHC and the FCC.
%
As expected, increasing $m_{hh}^\text{max}$, {\it i.e.}
making the cut less stringent,
leads to stronger constraints.
Eventually, $\dcvv$ flattens out when $m_{hh}^\text{max}$
is large enough that all the $m_{hh}$ bins
which contain at least one event are included in the
posterior probability of Eq.~(\ref{eq:postpdf1}).
Bins which contain at least one event are depicted in Fig.~\ref{fig:c2v-sens} with a solid curve, while those containing less than one event are depicted by a dashed curve.
The gray area in Fig.~\ref{fig:c2v-sens} corresponds to the non-perturbative
region where $\dcvv > \dcvv^\text{max}$, obtained by setting 
$\Lambda = m_{hh}^\text{max}$ in Eq.~(\ref{eq:dcvvmax}),
the most optimistic assumption compatible with the validity of the EFT expansion.

As an additional way to quantify the validity of the EFT approach in our analysis,
we derive the region of exclusion in the plane $(\Lambda, g_*)$~\cite{Contino:2016jqw},
corresponding to the limits on $\dcvv$ derived as a function of $m_{hh}^{\rm max}$.
This is shown in the left panel of Fig.~\ref{fig:c2v-sens},
making use of Eq.~(\ref{eq:NDA}) and then setting
$\Lambda = m_{hh}^\text{max}$.
The gray area in the upper part of the plot indicates the non-perturbative region, 
defined by $g_*\ge 4\pi$.
%
We find that the dominant constraints on $\dcvv$ arise from a region
in the parameter space where the EFT expansion is valid, both at the
LHC and at the FCC.
The results from the two panels of Fig.~\ref{fig:c2v-sens} indicate that
our EFT analysis is robust and that can be used to derive stringent bounds
on $\dcvv$ in the absence of new explicit degrees of freedom.

\begin{figure}[t]\centering
  \includegraphics[scale=1]{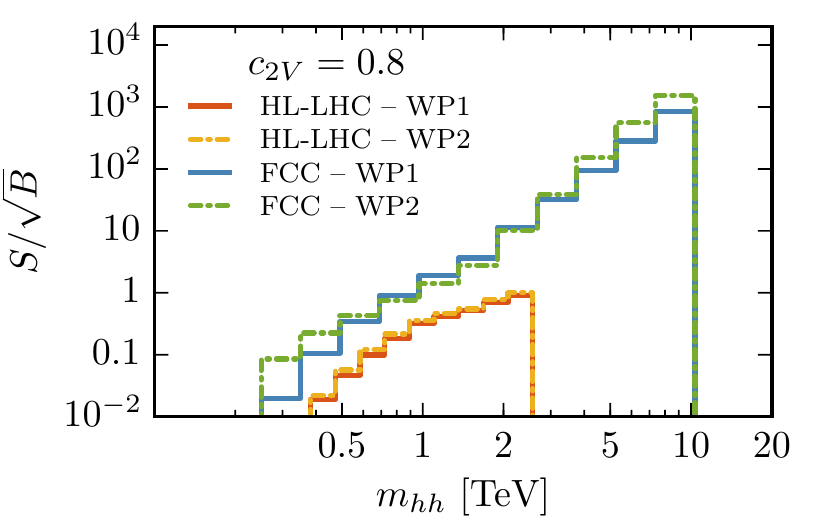}
	\caption{\small Signal significance $S/\sqrt{B}$
          in the $c_{2V}~$=$~0.8$ scenario, as a function of
          the di-Higgs invariant mass $m_{hh}$
          at the HL-LHC and the FCC, for the two $b$-tagging working
          points of Eq.~(\ref{eq:btaggingWP}).
         }
	\label{fig:significance}
\end{figure}

Finally, in Fig.~\ref{fig:significance} we show the signal significance, $S/\sqrt{B}$,
          in the $c_{2V}~$=$~0.8$ scenario as a function of
          the di-Higgs invariant mass $m_{hh}$
          at the HL-LHC and the FCC.
          The results are presented
          for the two $b$-tagging working
          points defined in Eq.~(\ref{eq:btaggingWP}). As already discussed, these have been chosen so that
the first (WP1) is consistent with the current ATLAS and CMS performances, while the second (WP2)
assumes an improved detector performance.
          %
          One can observe that the signal significance of each individual bin is at most
          $S/\sqrt{B}\simeq 2$
          at the HL-LHC (though the precise numbers depend on the specific choice of binning),
          while at the FCC one finds much higher signal significances,
          with $S/\sqrt{B}\simeq 5$ already for $m_{hh}\simeq 1.5$ TeV and then
          increasing very rapidly for higher values of $m_{hh}$.
          Figure~\ref{fig:significance} clearly shows that the signal significance
          depends very mildly on the specific details of the $b$-tagging performance and that
operating at WP2 instead of WP1 implies only a minor improvement.
          
To summarize, we have demonstrated how Higgs boson pair production
via VBF can be used to provide the first direct constraints
on the $c_{2V}$ coupling already at the LHC with
$\mathcal{L}=300$ fb$^{-1}$ (Table~\ref{tab:resultsdcvv}), which at
a $100\,$TeV collider would become a high-precision measurement
with potentially sub-percent accuracy.
We have also assessed (Fig.~\ref{fig:c2v-sens}) the robustness
of our strategy and the validity of the underlying
EFT expansion.
Our analysis clearly highlights the unique physics potential
of extending current di-Higgs searches at the LHC to the
vector-boson fusion channel.

%% file: sec-conclusions.tex
\section{Conclusions and Outlook}
\label{sec:conclusions}

The measurement and study of Higgs pair production is one of the cornerstones of
the LHC  program as well as of any future hadron collider. It provides unique
information on the Higgs sector and on the mechanism underlying electroweak
symmetry breaking, and allows a direct test of the strength of the Higgs boson
self-interactions. On the other hand, it is a challenging measurement and the
low production rates require large integrated luminosities to achieve reasonable
signal significances. While most studies of Higgs pair production so far have
concentrated on the gluon-fusion channel which has the largest cross section,
we have shown in this work how the vector-boson fusion channel can impose
stringent constraints on Higgs couplings that are not directly accessible by
other means, in particular on the $hhVV$ quartic coupling $c_{2V}$.

Exploiting the high signal yield of the $b\bar{b}b\bar{b}$ final state, we have 
presented a detailed feasibility study of the measurement of Higgs boson pairs
in the vector-boson fusion channel at the LHC and at a future 100 TeV hadron
collider. A key ingredient of our strategy is provided by the fact that
deviations of the Higgs couplings to vector bosons from the parabola
$\cvv=\cv^2$ significantly harden the $m_{hh}$ distribution resulting in a large
fraction of events with a boosted Higgs pair. The subsequent decays into
$b\bar{b}$ pairs can then be reconstructed by means of jet substructure
techniques. While QCD backgrounds are very large, we have shown how the
combination of selection cuts exploiting the  VBF topology and the growth of the
$m_{hh}$ distribution when the Higgs couplings depart from their SM values leads
to a remarkable model-independent sensitivity to the $c_{2V}$ coupling.

Our results demonstrate that at the LHC with an integrated luminosity of
$\mathcal{L}=300$ (3000) fb$^{-1}$ the $hhVV$ coupling can be measured with
$_{-37\%}^{+45\%}$ ($_{-15\%}^{+19\%}$) precision at the 68\% probability level,
reaching 1\% accuracy at a $100\,$TeV collider. Therefore, stringent constraints
on this so far unknown coupling can be obtained already before the start of the
HL-LHC data taking. Our analysis provides strong motivation for the ATLAS and
CMS collaborations to extend their searches for Higgs pair production to the VBF
channel already during Runs II and III. On the other hand, we also find that the
VBF channel is clearly inferior to the gluon-fusion channel for a measurement of
the Higgs self-coupling~$\lambda$, at least for the $b\bar{b}b\bar{b}$ final
state studied here, since the sensitivity to $\lambda$ arises from the threshold
region $m_{hh}\simeq 2m_h$ where QCD backgrounds overwhelm the signal even for
sizeable modifications of the Higgs couplings with respect to their SM values.

There are several possible avenues for future work. On one hand, it might be
interesting to study the possibility to enhance the sensitivity to $\cvv$ by
means of a multivariate analysis (MVA), such as those used
in~\cite{Behr:2015oqq}, in order to dynamically determine the optimal set of
selection cuts and optimise the discrimination between signal and background
events. Further, it should be possible to quantify the constraints on additional
EFT operators that can contribute to the di-Higgs VBF signal yield and that have
not been considered in this work. Finally,  a complete analysis should include a
full detector simulation, especially for the reconstruction of the forward jets
and of the Higgs boson candidates, and a $b$-tagging strategy able to reproduce
more closely the one adopted by the LHC experiments.

\noindent
\subsection*{Acknowledgements}

We thank Nathan Hartland for assistance with the {\tt Sherpa} event generation
and for many discussions about Higgs pair production.
F.~B. is supported a STFC Consolidated Grant  and acknowledges
the hospitality and support of the CERN Theory department. 
The work of R.~C. was partly supported by the ERC Advanced Grant No. 267985,
\textit{Electroweak Symmetry Breaking, Flavour and Dark Matter: One Solution for 
Three Mysteries (DaMeSyFla)}. 
J.~R. is supported  by the ERC Starting Grant ``PDF4BSM''.

%% file: sec-mc.tex
\section{Monte Carlo event generation}
\label{sec:mc}

In this Appendix we discuss the event generation of the signal and
background process used in this analysis.
In each case, we discuss the programs used, the input parameters
and theoretical settings, and the cross-checks that have been performed.

Signal events for Higgs pair production via VBF
have been generated at leading order (LO)
with \verb|MadGraph5| \verb|aMC@NLO|~\cite{Alwall:2014hca}
using a tailored {\sc\small UFO}~\cite{Degrande:2011ua} model that allows
for generic values of the  $c_V$, $c_{2V}$,
and $c_3$ couplings in the 
Lagrangian Eq.~(\ref{eq:lagrangian}) (see~\cite{Alloul:2013naa,Artoisenet:2013puc}
for other {\sc\small UFO}
models of Higgs EFTs).
We generated $1$M unweighted events for each value of $c_{2V}$
and for center-of-mass energies of 14 TeV and 100 TeV.
The size of the signal sample is dictated
by the condition of achieving an adequate
coverage of the large $m_{hh}$ region.
Even so, for small deviations of $\cvv$ with respect
to its SM value this region is very difficult to populate.
We overcome this limitation by performing a fit to the cross-section
in each $m_{hh}$ bin as a
function of $\cvv$ using the general parametrization of Eq.~(\ref{eq:xsecgeneral}).

\begin{table}\centering\renewcommand{\arraystretch}{1.3}
  \small
	\begin{tabular}{r @{\hskip 5mm} c c@{\hskip 1cm} c c}
		\toprule[1pt]
	&	\multicolumn{2}{c}{14 TeV}	&	\multicolumn{2}{c}{100 TeV}\\
									& Signal& QCD bkg.& Signal& QCD bkg.\\\midrule
$p_{T_j}~\text{(GeV)}~\geq~$&	25	&	20	&	25	&	20\\
$p_{T_b}~\text{(GeV)}~\geq~$&	25	&	20	&	25	&	20\\
$|\eta_j|\leq~$				&	4.5	&	3	&	10	&	3\\
$|\eta_b|\leq~$				&	2.5	&	3	&	3	&	3\\
$\Delta R_{bb}\geq~$		&	0.1	&	0.1	&	0.1	&	0.1\\
$\Delta R_{jb}\geq~$		&	0.4	&	0.1	&	0.4	&	0.1\\
$\Delta R_{jj}\geq~$		&	4	&	0.1	&	4	&	0.1\\
$m_{jj}~\text{(GeV)}~\geq~$	&	600	&	---	&	600	&	---\\
		\bottomrule[1pt]
	\end{tabular}
	\caption{\small Parton-level generation cuts 
          for the  VBF di-Higgs signal
          and the QCD background samples at $\sqrt{s}=14$ TeV and 100 TeV.
        In this table, $j$ refers to light quarks and gluons, $b$ to bottom quarks,
        and $m_{jj}$ is the invariant mass of the two light partons in the event
        with largest $p_T^j$.
}
	\label{tab:gen-cuts}
\end{table}

The  {\tt MadGraph5\_aMC@NLO}  generation of signal events uses
the NNPDF2.3LO~\cite{Ball:2012cx} set with $\alpha_S(m_Z)=0.119$
interfaced via {\tt LHAPDF6}~\cite{Buckley:2014ana}.
The calculation is performed in the
$n_f=4$ scheme and the factorization and renormalization
scales are taken to be the $W$ boson mass,
$\mu_F=\mu_R=M_W$.
To account for higher order effects, we apply an NNLO/LO $K$-factor $\simeq 1.1$
on the inclusive cross-section~\cite{Liu-Sheng:2014gxa} both at $14\,$TeV and
at $100\,$TeV.
The decays of the Higgs bosons into $b\bar{b}$ pairs are
performed within {\tt MadGraph5\_aMC@NLO} with adjusted
parameters
to ensure that the branching ratio corresponds
to Higgs Cross-Section Working
Group value~\cite{deFlorian:2016spz} of ${\rm BR}(h\to b\bar b)=0.582$.
Since in this work we only modify the $\cvv$ coupling,
assuming the SM value of the total Higgs width is
a very good approximation.

With this setup, we have generated signal events in the SM and for a
wide range
of values of $\dcvv$.
At the matrix-element level, we applied the generation cuts listed
in Table~\ref{tab:gen-cuts}.
The corresponding cross sections are summarized
in Table~\ref{tab:summarygen}, where we provide
the SM results as well as the predictions for various
BSM scenarios defined by
the couplings  $\{c_V,c_{2V},c_3\}$,
including the benchmark scenario $c_{2V}=0.8$.
In addition, the number of expected events 
at the HL-LHC, assuming an
integrated luminosity of $\mathcal{L}=3\,{\rm ab}^{-1}$, and
at the FCC 100 TeV for $\mathcal{L}=10\,{\rm ab}^{-1}$, are also listed.

\begin{table}[t]
  \centering
  \small
\renewcommand{\arraystretch}{1.3}
\begin{tabular}{c c c c c c}\toprule[1pt]
\multicolumn{6}{c}{ Signal: VBF $hh\to b\bar{b} b\bar{b}$} \\
\toprule[1pt]
\multirow{2}{*}{$\{c_V,c_{2V},c_3\}$} & & \multicolumn{2}{c}{LHC 14 TeV} & \multicolumn{2}{c}{FCC 100 TeV} \\
  &   & $\sigma$  (fb) & $N_{\rm ev}(\mathcal{L}=3\,{\rm ab}^{-1})$ & 
 $\sigma$  (fb) & $N_{\rm ev}(\mathcal{L}=10\,{\rm ab}^{-1})$ \\\midrule
\{1,1,1\} & SM        & 0.26 & 780 & 15 & $1.5\cdot 10^5$ \\
\hline
\{1,0,1\}   &          & 4.4 & $1.3\cdot 10^4$ & 593 & $5.9\cdot 10^6$ \\
\{1,2,1\}    &        & 2.5 & $7.5\cdot 10^3$ & 471 & $4.7\cdot 10^6$ \\
\{1,0,0\}     &        & 5.8 & $1.7\cdot 10^4$ & 656 & $6.6\cdot 10^6$ \\
\{1,0,-1\}  &          & 7.5 & $2.3\cdot 10^4$ & 731 & $7.3\cdot 10^6$ \\
\{1,1,0\}    &         & 0.64 & $1.9\cdot 10^3$ & 30 & $3.0\cdot 10^5$ \\
\hline
\{1,0.8,1\}    &  Benchmark & 0.58 & 1740 & 48 & $4.8\cdot 10^5$ \\
\bottomrule[1pt]
\end{tabular}
\caption{\label{tab:summarygen}
  Parton-level cross sections for VBF Higgs pair production
   in the $b\bar{b}b\bar{b}$ final
   state at 14 TeV and  100 TeV (including 
   branching ratios)
   after the acceptance cuts of Table~\ref{tab:gen-cuts}.
   We show the cross-sections
   for different values of the couplings  $\{c_V,c_{2V},c_3\}$
   normalized to the SM values.
We also provide the number of events at the HL-LHC for
$\mathcal{L}=3\,{\rm ab}^{-1}$ and at the FCC for
$\mathcal{L}=10\,{\rm ab}^{-1}$.
}
\end{table}

Following the parton-level event generation, the resulting Les Houches
event files for signal events
are showered using the {\tt Pythia8}
Monte Carlo generator~\cite{Sjostrand:2014zea}
v8.212 using the Monash 2013 Tune~\cite{Skands:2014pea}.
No hadronization or underlying event (UE) effects are included for simplicity.
Although we have not attempted to simulate the effects of pile-up (PU) in our analysis,
recent studies~\cite{Behr:2015oqq} indicate that modern PU
mitigation techniques~\cite{Cacciari:2014gra} should be efficient
enough to minimize the PU contamination even for the $b\bar{b}b\bar{b}$
final state.

Concerning the generation of the QCD background processes,
their parton-level cross sections
are summarized in Table~\ref{tab:summarygenBack}.
In each case, we indicate the programs used for the event generation,
the number of MC events $N_{\rm ev}$ that have been generated,
 whether the events are weighted or unweighted,
and the LO cross sections along with the corresponding
higher-order $K$-factors at 14 TeV
and 100 TeV.
As discussed in Sec.~\ref{sec:results}, our results
include a
15\% systematic uncertainty on the total background normalization
(assuming a data-driven determination
in an experimental analysis), and we also
assessed the robustness of
our estimate for the measurement of $\dcvv$ in case of an overall
upwards shift of the backgrounds by a factor 3.

For QCD multijet production we have explored two complementary approaches for
event generation. First of all, we used {\tt ALPGEN}~\cite{Mangano:2002ea} to
generate a large sample of unweighted events at the matrix-element level and
showered them with {\tt Pythia8}. This approach, however, presents a difficulty,
because the differential cross section with respect to the di-Higgs invariant
mass $m_{hh}$ falls very rapidly for increasing $m_{hh}$. We thus finds that an
unrealistically large sample of unweighted events would be needed in order to
adequately populate the tail of the $m_{hh}$ distribution.

\begin{table}[t]
  \centering
  \footnotesize
\renewcommand{\arraystretch}{1.3}
\begin{tabular}{c c c c c c c c}\toprule[1pt]
\multicolumn{7}{c}{ Background processes} \\[0.3pt]
\toprule[1pt]
  \multirow{2}{*}{Process} & \multirow{2}{*}{Program}  & \multirow{2}{*}{$N_{\rm ev}$}  &
  \multicolumn{2}{c}{$\sigma_{\rm LO}$ (fb)} &
  \multicolumn{3}{c}{$K$--factor}   \\
  & & & LHC14 & FCC100& LHC14 & FCC100  & Ref.\\\midrule
  $4b$  & {\tt Sherpa2.2} & 50M [W]  & $1.1\cdot 10^6$  &
  $1.6\cdot 10^7$ & 1.7 & 1.7 &\cite{Alwall:2014hca}\\
  $2b2j$  & {\tt Sherpa2.2} & 50M [W]  & $2.6\cdot 10^8$  &
  $3.8\cdot 10^9$&  1.3 & 1.3 &\cite{Alwall:2014hca}\\
  $t\bar{t}jj$  & {\tt Sherpa2.2} & 10M [W]  & $1.9\cdot 10^4$ &
  $1.6\cdot 10^6$ & 1.6 & 1.6 &\cite{Czakon:2013goa}\\
  \hline
  $4b2j$  & {\tt ALPGEN}   & 6M(2M) [UW]  & $5.4\cdot10^{4}$ & $2.4\cdot10^{6}$  & 1.7 & 1.7 & --\\
  $2b4j$  & {\tt ALPGEN}   & 260k [UW]  & $10^{7}$ & $5.2\cdot10^{8}$ &  1.3 & 1.3  &--\\
  \hline
  $gg\to hh\to b\bar{b} b\bar{b}$  & {\tt aMC@NLO} &
  1M [UW] & 6.2  & 272 & 2.4  & 2.2 & \cite{deFlorian:2015moa}\\
\bottomrule[1pt]
\end{tabular}
\caption{\label{tab:summarygenBack} \small
Parton level cross-sections for the background processes considered in
this work after the acceptance cuts of Table~\ref{tab:gen-cuts}.
We indicate the programs used, the numbers of events generated $N_{\rm ev}$ and
whether they are weighted [W] or unweighted [UW],  the LO cross sections and the
corresponding $K$-factors and the relevant reference. In the case of the $4b2j$
and $2b4j$ backgrounds, the same $K$-factors as those for the $4b$ and $2b2j$,
respectively, were applied. For the $4b2j$ sample, the 6M(2M) events correspond to those generated at the LHC(FCC) centre of mass energies, respectively. The generation cuts used for the \alpg~samples are
different, see text.
}
\end{table}

To bypass this limitation, we generated 50M weighted $4b$ and
$2b2j$ events with {\tt Sherpa}
v2.2~\cite{Gleisberg:2008ta} and then processed them with the built-in shower.
The main advantage of this approach is that events with small weights are
not discarded by the unweighting and, therefore, enough events with
large $m_{hh}$ (and thus small weight) are still kept.
One possible drawback is that, for the same number of events,
the unweighted sample provides a better estimate of the total cross section
than the weighted one.
In our case this is not an issue, since with
our weighted sample we achieve statistical uncertainties of order
$2\%$ (to be compared with $\gsim 0.01\%$ for a same-size
unweighted sample), which is more than sufficient for our purposes.
On the other hand, to generate a large enough number of events in a reasonable amount of CPU time 
we are forced to use a lower multiplicity final state,  and thus we rely  on
the parton shower to generate the additional light partons
(see Table~\ref{tab:summarygenBack}).
To validate this procedure, we explicitly
verified that, in the region where the
two approaches
lead to sufficient statistics, the {\tt Sherpa}
calculation based on $4b$ ($2b2j$) matrix elements and
the {\tt ALPGEN} one, based on
$4b2j$ $(2b4j)$ matrix elements, lead to comparable results
(see~\ref{sec:qcdmultijet}).
This agreement indicates
that the parton shower does a reasonable job in modelling
additional hard radiation in the phase-space region of interest.

The {\tt Sherpa} samples in Table~\ref{tab:summarygenBack}
have been generated using the 
NNPDF3.0 NNLO set~\cite{Ball:2014uwa}
with strong coupling
 $\alpha_S(m_Z^2)=0.118$ and with $n_f = 4$ active quark flavours,
 and use as
 factorization and renormalization scales $\mu_F=\mu_R=H_T/2$,
 with
\begin{equation}
H_T \equiv \sum_{i}^{n}\sqrt{\lp m_{t,i}\rp^2+\lp {p_{T,i}^t}\rp^2} \, ,
\end{equation}
where $n$ is the final-state matrix-element multiplicity and
$m_{t,i}$ and $p_{T,i}^t$ are the transverse mass
and transverse momentum of the $i$-th final-state parton.
The generation-level acceptance cuts applied to these samples are listed
in Table~\ref{tab:gen-cuts}.
As for the other background samples, LO cross sections have been rescaled
by the best available higher-order $K$-factors.

Higgs pair production via gluon-fusion is simulated at LO using
{\tt MadGraph5\_aMC@NLO} for loop-induced processes~\cite{Hirschi:2015iia}.
The cross-section is rescaled to match the
inclusive NNLO+NNLL calculation~\cite{deFlorian:2015moa} 
yielding a $K$-factor, $\sigma_{\rm NNLO+NNLL}/\sigma_{\rm LO}=2.4~(2.2)$
at $14\,$TeV ($100\,$TeV).
Parton-level events are then showered with  {\tt Pythia8}
using the same settings as for the signal VBF di-Higgs samples.
No generation cuts are applied,
and the resulting cross sections (including
the branching fraction into $b\bar{b}b\bar{b}$)
are listed in Table~\ref{tab:summarygenBack}.
While the hard-scattering process that is generated, $gg\to hh$,
does not include additional jets,
initial state radiation from the gluon legs 
can give a VBF-like topology with two forward jets characterized by a large
invariant mass, and thus contribute to the total $hhjj$ yield.
We have verified that our calculation
provides a reasonable description of the relevant kinematical distributions,
such as $m_{hh}$, by comparing with the $gg\to hhjj$ process computed in the EFT approximation.
As shown in Sec.~\ref{sec:results}, while the gluon-fusion
contamination to the VBF signal is substantial close to threshold,
it is marginal in the large $m_{hh}$ region where
the sensitivity to $c_{2V}$ is the highest.

%% file: sec-fitting.tex
\section{Fitting the tail of the $m_{hh}$ distribution for background processes}
\label{sec:fitting}

The background processes considered in this work (see~\ref{sec:mc})
exhibit a steep fall-off for
large values of  $m_{hh}$, the invariant mass distribution of the reconstructed Higgs
pair.
Even with the use of weighted events, it is
difficult to adequately populate this region.
To obtain a
reliable estimate of the cross section there, it is thus necessary to introduce a
fitting procedure.
In this Appendix we discuss how the fits
to the $m_{hh}$ distributions for the background processes, and the 
associated validation tests, have been performed.

We found that, far from the di-Higgs production threshold,
the following functional form provides a reasonable
description of the $m_{hh}$ distribution:
\begin{equation}
  \label{eq:mhhfit}
\sigma^{\rm (fit)}(m_{hh},\sqrt{s}) = \lc 1-\lp \frac{m_{hh}}{\sqrt{s}}\rp^{1/4}\rc ^a\lp \frac{m_{hh}}{\sqrt{s}}\rp^{b+c\cdot\log (m_{hh}/\sqrt{s})}\, ,
\end{equation}
where $\sqrt{s}$ is the collider centre-of-mass energy.
This specific functional
form is a modified version of one of those suggested
in Ref.~\cite{Davis:2016hlw} for the fit of the di-photon invariant mass
distribution at the LHC.
We explored other choices, finding a comparable
fit quality.

In Eq.~(\ref{eq:mhhfit}), the fit parameters $\{a,b,c\}$ are determined
by minimizing the $\chi^2$,
\begin{equation}
  \label{eq:chi2}
\chi^2 = \sum_{i=1}^n \frac{\lp \sigma^{\rm (th)}_i-\sigma^{\rm (fit)}(m_{hh}^{(i)},\sqrt{s})\rp^2}{\delta_i^2} \, ,
\end{equation}
where $n$ is the number of bins in the $m_{hh}$ distribution used as input to
the fit, $\sigma^{\rm (th)}_i$ is the theoretical prediction for the
cross-section in the bin $m_{hh}^{(i)}$, and
\begin{equation}
\label{eq:stat}
\delta_i \equiv
\sqrt{\sum_{k=1}^{N_i}\lp w_k^{(i)}\rp^2} \, ,
\end{equation}
is the statistical uncertainty associated to the
$N_i$ weighted Monte Carlo events that populate bin $i$ with weights $\{w_k^{(i)}\}$.
Note
that when the MC events are unweighted, $\delta_i\propto\sqrt{N_i}$, as
expected.
The resulting fit coefficients are shown in
Table~\ref{tab:bkg-coeffs} for both
centre-of-mass energies and for the four background
processes considered.
We have also verified that the results for the fit parameters are stable with respect to variations
of the binning in $m_{hh}$.

\begin{table}[t]\centering
	\renewcommand{\arraystretch}{1.3}
	\begin{tabular}{c c c c c}\toprule[1pt]
		$\sqrt{s}$ [TeV] & Background & $a$ & $b$ & $c$\\\midrule
		\multirow{4}{*}{14}
		& $4b$ 		& 14.2 	&  0.492	&  0.512\\
		& $2b2j$ 	& 24.9 	& -2.65 	&  0.198\\
		& $ttjj$ 	& 48	& -18.3		& -3.63	\\
		& $ggf$ 	& 30.1 	& -8.27		& -1.82	\\
		\midrule
		\multirow{4}{*}{100}
		& $4b$ 		& 32.3 	& -4.82	& -0.539	\\
		& $2b2j$ 	& 36.8 	& -2.97	& -0.118	\\
		& $ttjj$ 	& 38.4 	& -3.31	& -0.265	\\
		& $gg\to hh$& 15.9  &  3.61	&  0.695 	\\
		\bottomrule[1pt]
	\end{tabular}
	\caption{\small The fit
          coefficients in Eq.~(\ref{eq:mhhfit}) for each background process
          at $14$ TeV and 100 TeV.
	}
	\label{tab:bkg-coeffs}
\end{table}

\begin{figure}[h]
	\centering
	\begin{minipage}{0.49\textwidth}\centering
		\hspace{1cm}LHC -- 14 [TeV]
		\includegraphics[width=\textwidth]{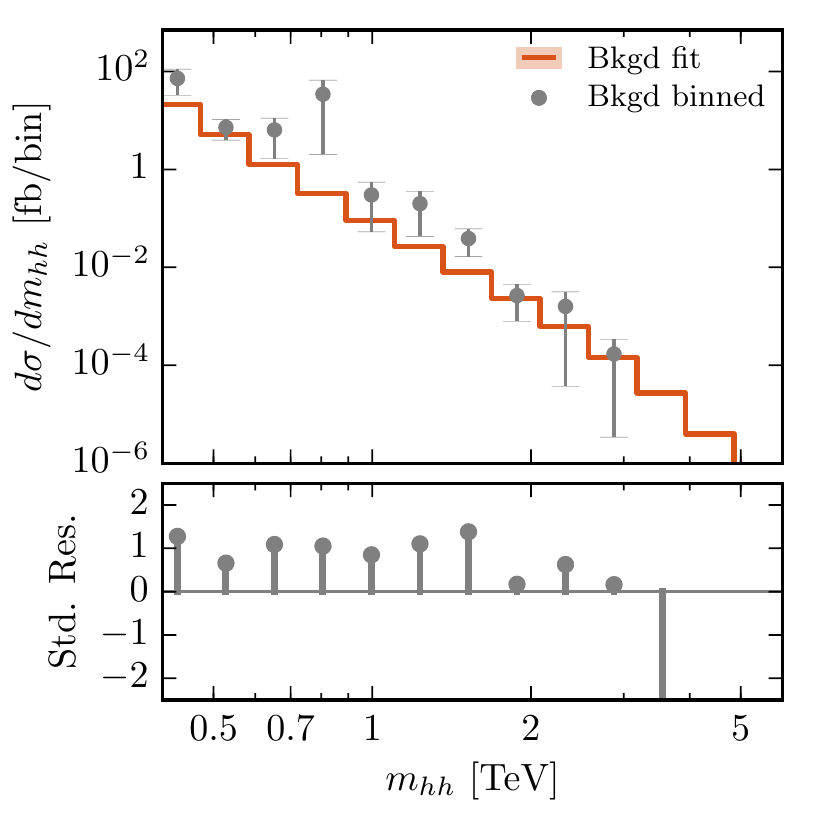}
	\end{minipage}
	\begin{minipage}{0.49\textwidth}\centering
		\hspace{1cm}FCC -- 100 [TeV]
		\includegraphics[width=\textwidth]{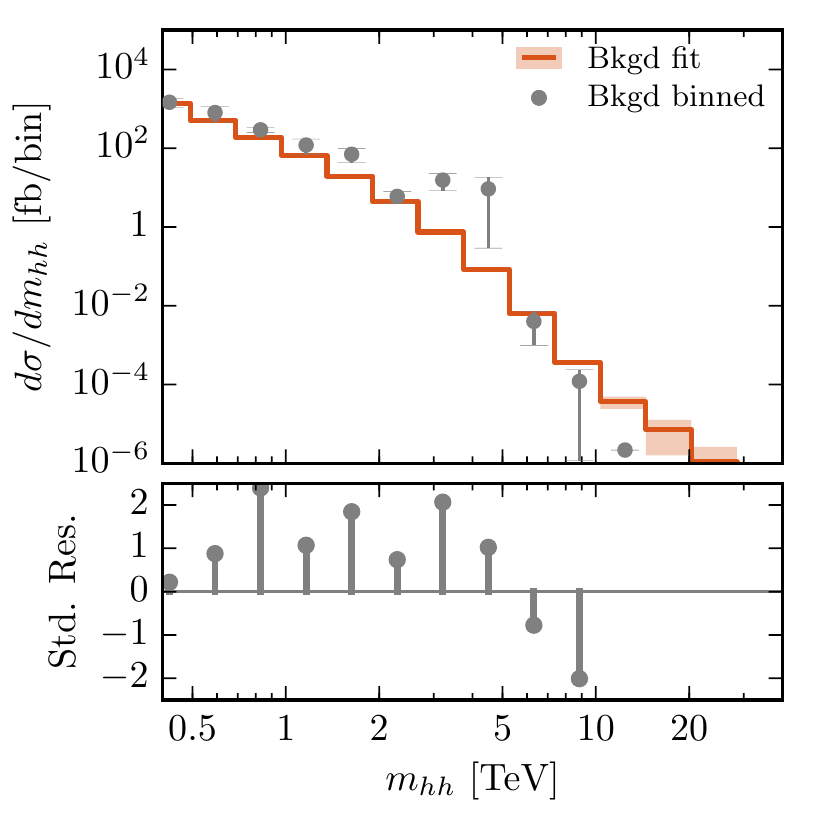}
	\end{minipage}
	\caption{\small
Invariant mass distribution of the di-Higgs system
for the total background at $14\,$TeV (left) and $100\,$TeV (right).
The histograms are obtained by summing the contributions from
the separate fits to each background processes.
We also show the binned MC events (filled circles) with their
statistical uncertainty.
The lower panels show the standardized residuals in each bin. 
}
\label{fig:bkg-fits}
\end{figure}

To validate the fitting procedure, we show in Fig.~\ref{fig:bkg-fits} the
$m_{hh}$ distribution of the reconstructed di-Higgs system
for the total background, where
each individual process has been fitted separately and
then added up to construct the solid
histograms.
The fit results are compared to the cross-sections from the
Monte Carlo generation (indicated by filled circles) with their corresponding
statistical uncertainty, Eq.~(\ref{eq:stat}).
In the lower panels of Fig.~\ref{fig:bkg-fits} we show the fit
residuals, defined as the difference between MC and fit divided
by the statistical uncertainty in each bin.
The fact that the parametrized fit agrees
with the MC calculations at the 2-$\sigma$ level or better (in units of the
uncertainty of the latter) for a wide range of $m_{hh}$ 
demonstrate the goodness of these fits.

Note that, in the case of the QCD multijet backgrounds, we exclude
the first bin to ensure that the fit is not affected by
artificial features in the $m_{hh}$ distribution
induced by the analysis selection cuts.
Furthermore, in the
case of the gluon fusion di-Higgs background, it is necessary to exclude the first few
bins of the $m_{hh}$ distribution from the fit.
The reason is that in this case there is a
production threshold at $2m_{h}$, and, as a result, the distribution does
not decrease monotonically with $m_{hh}$ unless one is far enough from threshold.
By excluding these bins, we
avoid biasing the resulting fit in the tail of the $m_{hh}$ distribution,
the region where a functional form such as Eq.~(\ref{eq:mhhfit})
does provide an equally satisfactory description as for the rest of background processes.


%% file: sec-qcdmultijet.tex
\section{Validation of the QCD multijet generation}
\label{sec:qcdmultijet}

As discussed in~\ref{sec:mc}, the modelling of
the  tail of
the  $m_{hh}$ distribution for background processes
is particularly challenging.
One reason is because
all the backgrounds considered
are characterized by a steep power-like fall-off with
$m_{hh}$ above the di-Higgs production threshold,
and therefore it becomes necessary to introduce a
cross-section parametrization (see~\ref{sec:fitting})
to be able to cover this region.

In addition, in the case of the QCD
multijet backgrounds, there are different
options available for the modelling of the $m_{hh}$
distribution, in particular the
multiplicity of the matrix-element calculation
prior to the parton shower.
Ideally, one should
generate all relevant final-state partonic
multiplicities and merge them to  avoid double counting, either at
LO~\cite{Mangano:2001xp,Catani:2001cc,Alwall:2007fs}
or at NLO~\cite{Hoeche:2012yf,Frederix:2012ps}.
For the purposes of this feasibility study, however,
we found it sufficient to generate
LO samples using $4b$ and $2b2j$ matrix elements with
{\tt Sherpa}, with additional hard radiation provided
by the parton shower.
Note that our approach  could
introduce some double counting from gluon splittings into $b\bar b$, which if
anything would increase the background cross section and make
our estimates of $\dcvv$ more conservative.
Moreover, let us recall that, as discussed in Sec.~\ref{sec:results},
an actual experimental analysis
would estimate the overall background normalization
by means of data-driven techniques.

To validate the robustness of our simulation of the
$m_{hh}$ distributions for the QCD multijet background, we have
compared the {\tt Sherpa} calculation, based on $4b$ and $2b2j$ matrix elements
and weighted events, with the {\tt ALPGEN} simulation,
based on  $4b2j$ and $2b4j$ matrix elements and unweighted events 
showered with {\tt Pythia8}.
In principle, {\tt ALPGEN} should provide a better description
of the kinematics of the VBF jets, since it is based
on higher-multiplicity matrix elements, but it has the drawback that
populating the tail of the  $m_{hh}$ with unweighted events is
very CPU-time intensive.
On the other hand, it turns out that
the two approaches give comparable results for the $m_{hh}$ distribution
in the region where both approaches lead to
sufficient MC statistics, validating the use of the
{\tt Sherpa} for the calculation of background cross sections.

\begin{figure}[t]
	\centering
	\begin{minipage}{0.49\textwidth}\centering
		\hspace{1cm}LHC 14 TeV
		\includegraphics[width=\textwidth]{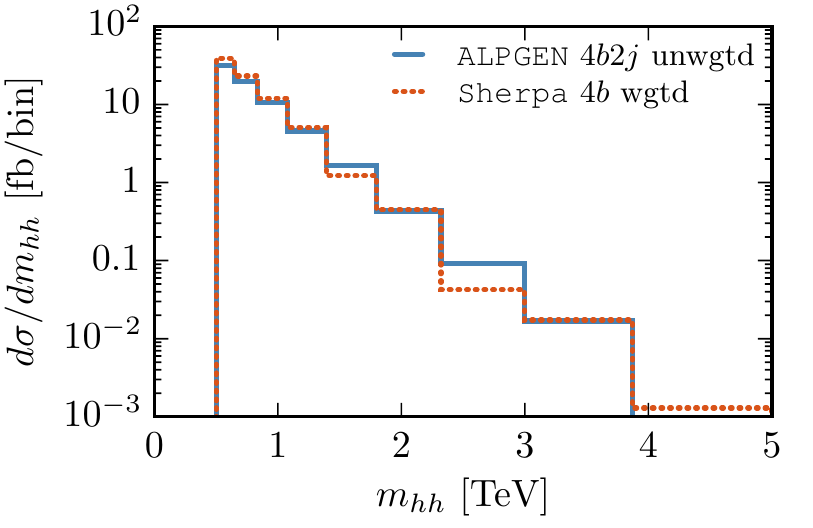}
	\end{minipage}
	\begin{minipage}{0.49\textwidth}\centering
		\hspace{1cm}FCC 100 TeV
		\includegraphics[width=\textwidth]{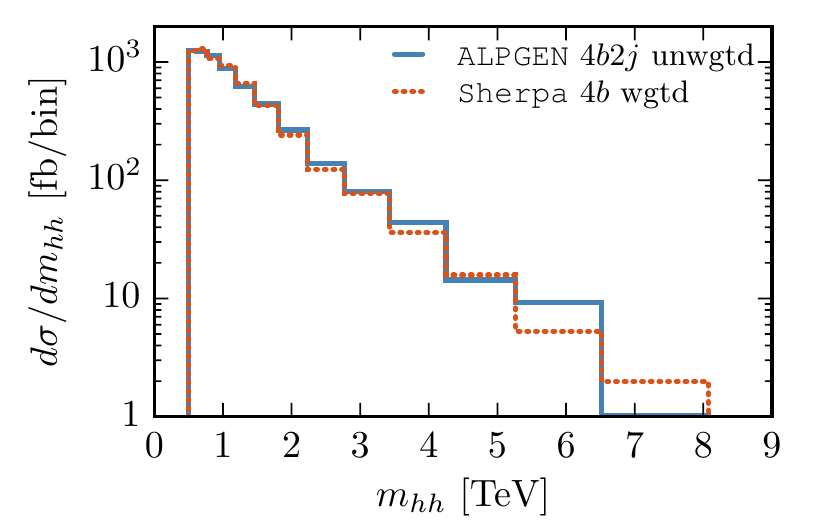}
	\end{minipage}
	\caption{\small \label{fig:bkg-comparison}
          Di-Higgs invariant mass distribution
          for the $4b$ background, comparing the results of {\tt
	    ALPGEN}, where unweighted events
          are generated with $4b2j$ matrix elements,
          with those of {\tt SHERPA}, where weighted
          events are generated with $4b$ matrix elements,
at 14 TeV (left) and 100 TeV (right).	}
\end{figure}

In Fig.~\ref{fig:bkg-comparison} we show the $m_{hh}$ distribution
for the $4b$ background at $14\,$TeV and $100\,$TeV,
comparing the {\tt ALPGEN} and {\tt SHERPA} calculations.
We find good agreement for the entire $m_{hh}$ range,
indicating that the {\tt SHERPA} event generation
does a reasonable job in modelling the VBF tagging jets,
and demonstrating that it can be reliably used to
populate the large $m_{hh}$
region for the multijet backgrounds with high efficiency.
The differences between the two calculations are at most a factor two,
typically less, well within the typical size of the theoretical
uncertainties for LO multijet calculations.

\begin{figure}[t]
	\centering
	\begin{minipage}{0.49\textwidth}\centering
          \hspace{1cm}LHC 14 TeV
          \includegraphics[width=\textwidth]{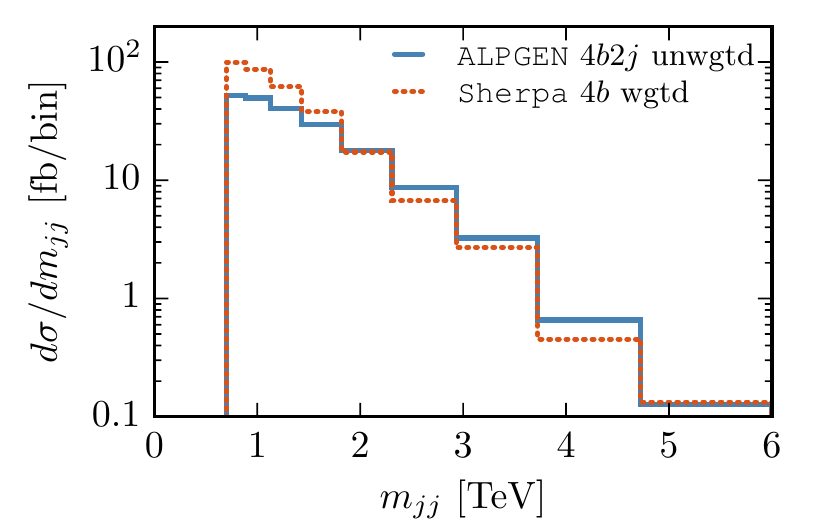}
          \includegraphics[width=\textwidth]{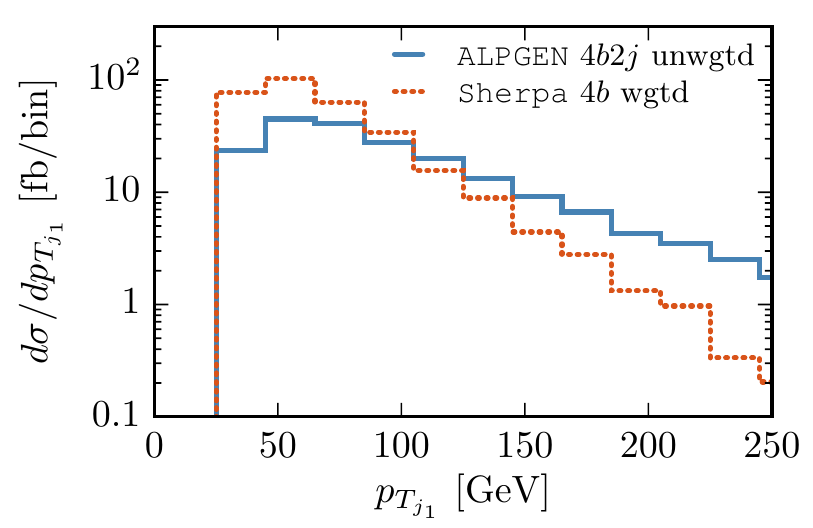}
       	\end{minipage}
	\begin{minipage}{0.49\textwidth}\centering
	  \hspace{1cm}FCC 100 TeV
           \includegraphics[width=\textwidth]{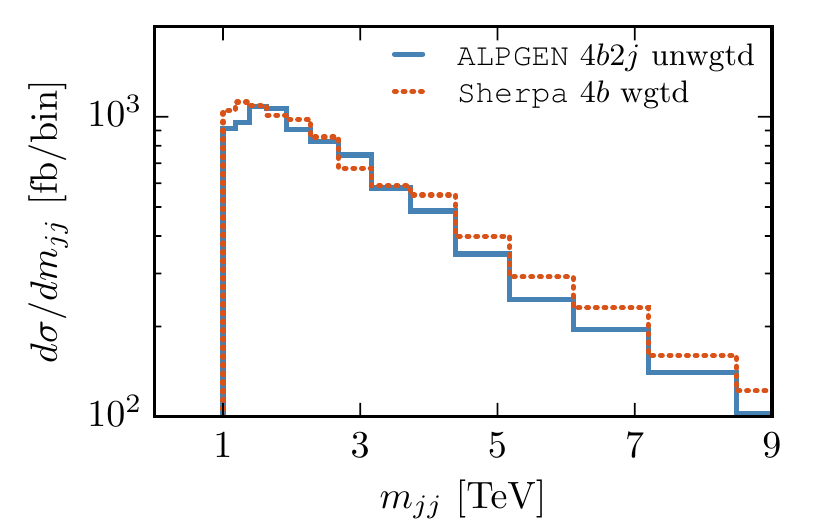}
           \includegraphics[width=\textwidth]{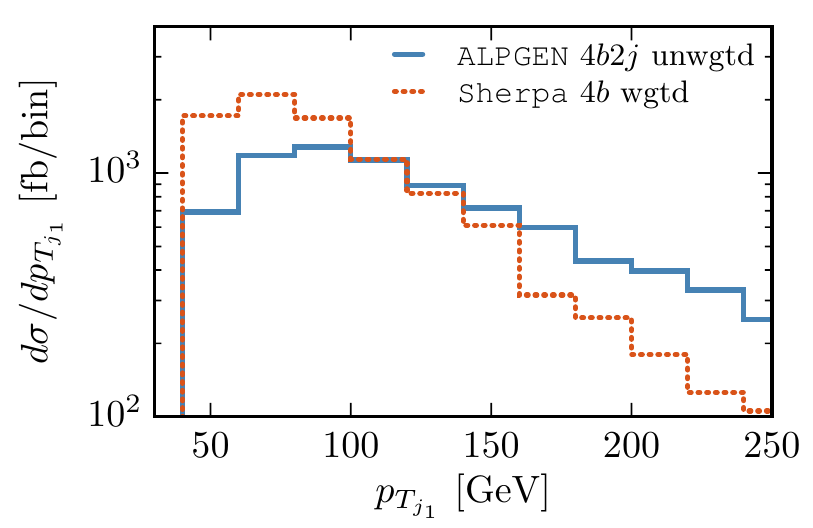}
	\end{minipage}
	\caption{\small \label{fig:bkg-comp2} 
          Same as Fig.~\ref{fig:bkg-comparison}, now 
          for the invariant mass of the two VBF
          tagging jets $m_{jj}$ (upper plots)
          and the transverse momentum
          of the hardest light jet in the event $p_{Tj_1}$ (upper plots).
	}
\end{figure}

Figure~\ref{fig:bkg-comp2}  shows a similar comparison for the invariant mass of the
two VBF tagging jets, $m_{jj}$, and  for the transverse momentum of the
hardest light jet in the event, $p_{Tj_1}$.
For the $m_{jj}$ distribution, the agreement between the
{\tt ALPGEN} and {\tt SHERPA} calculations is also good for the
entire kinematical range.
On the
other hand, for the transverse momentum of the leading jet
the {\tt ALPGEN} calculation leads to  harder
(softer) spectra than the {\tt SHERPA} one at high (low) values of $p_{Tj_1}$ .
This can be understood from the fact that $j_1$ will be typically generated
by the parton shower in the latter case, and by the matrix element in the former.
These differences are however inconsequential for our analysis, since the
cuts imposed on the $p_T$ of the VBF tagging jets are relatively mild,
see Table~\ref{tab:sel-cuts}, and thus the event selection will
not be affected.

The validation studies discussed in this Appendix demonstrate
that, for the purposes of the present analysis,
our approach to event generation based on
{\tt SHERPA} for the simulation of
the QCD multijet backgrounds is justified.
On the other hand, they also highlight that
future studies aiming to enhance the separation between
signal and background events from shape comparisons
of kinematical distributions, such as
multivariate analysis~\cite{Behr:2015oqq}, would require an
improved modeling of the QCD multijet backgrounds.
This could be achieved by using merging techniques
to combine QCD jet samples
of different multiplicities.

%% file: sec-fit-coeffs.tex
\section{Coefficients of the $\dcvv$ fit}
\label{sec:fit-coeffs}

The dependence of the signal cross section on $\dcvv$, as parametrized in
Eq.~(\ref{eq:xsecgeneral}), is required in order to construct the likelihood
function. In this appendix, we list the coefficients of
Eq.~(\ref{eq:xsecgeneral}) in each $m_{hh}$ bin. The coefficients are extracted
by fitting MC events after all cuts have been applied as discussed in
Sec.~\ref{sec:analysis} with the exception of the $m_{hh}$ cut. We use 15
equally spaced bins on a log scale starting from 250 GeV up to 6(30) TeV for
$\sqrt{s}=14(100)$ TeV. In addition, we define an overflow bin up to the centre
of mass energy. The results are shown in Tabs.~\ref{tab:fit-coeffs-14}~and~\ref{tab:fit-coeffs-100}. The fit error on each coefficient is also provided along with the off-diagonal entries of the correlation matrix $\rho_{ij}$ where $i,j\in\{\sigma,A,B\}$.

\begin{table}[t]\centering
	\setlength\tabcolsep{3pt}
	\renewcommand{\arraystretch}{1.3}
	\begin{tabular}{c c c c c c c}\toprule[1pt]
		Bin & $\sigma_\text{\sc sm}$ [fb] & A & B & $\rho_{\sigma A}$ & $\rho_{\sigma B}$ & $\rho_{AB}$\\\midrule
1 & $(3.08\pm0.05)\times 10^{-4}$ & $-3.61 \pm 0.0957$ & $6.89 \pm 0.217$ &		 $0.3$ & $-0.62$ & $-0.66$\\
2 & $(5.84\pm0.0691)\times 10^{-4}$ & $-3.96 \pm 0.0706$ & $7.27 \pm 0.161$ &		 $0.32$ & $-0.62$ & $-0.68$\\
3 & $(7.17\pm0.0654)\times 10^{-4}$ & $-4.3 \pm 0.0573$ & $9.71 \pm 0.148$ &		 $0.32$ & $-0.68$ & $-0.63$\\
4 & $(7.31\pm0.049)\times 10^{-4}$ & $-4.96 \pm 0.0462$ & $14.9 \pm 0.146$ &		 $0.35$ & $-0.77$ & $-0.59$\\
5 & $(5.98\pm0.0507)\times 10^{-4}$ & $-6.39 \pm 0.0677$ & $26.7 \pm 0.287$ &		 $0.42$ & $-0.86$ & $-0.59$\\
6 & $(4.19\pm0.0677)\times 10^{-4}$ & $-8.28 \pm 0.157$ & $50.1 \pm 0.93$ &		 $0.5$ & $-0.92$ & $-0.6$\\
7 & $(2.38\pm0.0571)\times 10^{-4}$ & $-11.9 \pm 0.302$ & $103 \pm 2.65$ &		 $0.61$ & $-0.96$ & $-0.67$\\
8 & $(1.15\pm0.0447)\times 10^{-4}$ & $-15.1 \pm 0.597$ & $174 \pm 7.05$ &		 $0.67$ & $-0.98$ & $-0.71$\\
9 & $(4.93\pm0.176)\times 10^{-5}$ & $-19.4 \pm 0.712$ & $301 \pm 11$ &		 $0.72$ & $-0.99$ & $-0.75$\\
10 & $(1.68\pm0.148)\times 10^{-5}$ & $-29.1 \pm 2.65$ & $758 \pm 67.8$ &		 $0.8$ & $-1$ & $-0.81$\\
11 & $(5.41\pm1.02)\times 10^{-6}$ & $-42.3 \pm 8.58$ & $1.69\times 10^{3} \pm 320$ &		 $0.83$ & $-1$ & $-0.83$\\
12 & $(1.28\pm0.506)\times 10^{-6}$ & $-47.4 \pm 22.6$ & $(3.88\pm1.53)\times 10^{3}$ &		 $0.76$ & $-1$ & $-0.77$\\
13 & $(8.55\pm13.4)\times 10^{-7}$ & $-16.5 \pm 29.7$ & $(2.27\pm3.57)\times 10^{3}$ &		 $0.76$ & $-1$ & $-0.77$\\
14 & $(3.5\pm3.57)\times 10^{-7}$ & $-0.00901 \pm 26.4$ & $(1.28\pm1.31)\times 10^{3}$ &		 $-0.064$ & $-1$ & $0.055$\\
15 & $(6.24\pm3.38)\times 10^{-7}$ & $-3.59 \pm 5.3$ & $105 \pm 60.7$ &		 $0.19$ & $-0.98$ & $-0.24$\\
16 & $(6.33\pm1.97)\times 10^{-7}$ & $-2.88 \pm 1.57$ & $22.1 \pm 8.59$ &		 $0.15$ & $-0.9$ & $-0.36$\\
		\bottomrule[1pt]
	\end{tabular}
	\caption{The bin by bin fit coefficients obtained by fitting MC events to Eq.~(\ref{eq:xsecgeneral}) for the LHC with $\sqrt{s}=14$ TeV. The first column labelled `Bin' gives the bin number in question. The bin definition is given in the text, see also footnote~\ref{fn:bin-def}. The last three columns labelled $\rho_{0A}$, $\rho_{0B}$, and $\rho_{AB}$ give the coefficients of the correlation matrix among the three fit parameters.}
	\label{tab:fit-coeffs-14}
\end{table}

\begin{table}[t]\centering
	\setlength\tabcolsep{3pt}
	\renewcommand{\arraystretch}{1.3}
	\begin{tabular}{c c c c c c c}\toprule[1pt]
		Bin & $\sigma_\text{\sc sm}$ [fb] & A & B & $\rho_{\sigma A}$ & $\rho_{\sigma B}$ & $\rho_{AB}$\\\midrule
1 & $(5.98\pm0.152)\times 10^{-3}$ & $-2.87 \pm 0.191$ & $7.8 \pm 0.513$ &		 $0.14$ & $-0.57$ & $-0.35$\\
2 & $(1.43\pm0.0185)\times 10^{-2}$ & $-4.29 \pm 0.106$ & $11.8 \pm 0.317$ &		 $0.23$ & $-0.62$ & $-0.46$\\
3 & $(2.25\pm0.0246)\times 10^{-2}$ & $-5.63 \pm 0.0999$ & $21.9 \pm 0.395$ &		 $0.29$ & $-0.74$ & $-0.46$\\
4 & $(2.36\pm0.045)\times 10^{-2}$ & $-7.95 \pm 0.212$ & $50.8 \pm 1.27$ &		 $0.37$ & $-0.86$ & $-0.47$\\
5 & $(1.88\pm0.0277)\times 10^{-2}$ & $-11.7 \pm 0.203$ & $109 \pm 1.86$ &		 $0.5$ & $-0.93$ & $-0.56$\\
6 & $(1.14\pm0.0226)\times 10^{-2}$ & $-14.6 \pm 0.339$ & $210 \pm 4.5$ &		 $0.56$ & $-0.97$ & $-0.6$\\
7 & $(5.68\pm0.196)\times 10^{-3}$ & $-24.9 \pm 0.985$ & $729 \pm 25.8$ &		 $0.69$ & $-0.99$ & $-0.7$\\
8 & $(2.73\pm0.224)\times 10^{-3}$ & $-43.2 \pm 4.07$ & $2.05\times 10^{3} \pm 170$ &		 $0.77$ & $-1$ & $-0.78$\\
9 & $(1.1\pm0.104)\times 10^{-3}$ & $-61.3 \pm 7.08$ & $5.54\times 10^{3} \pm 523$ &		 $0.77$ & $-1$ & $-0.77$\\
10 & $(2.97\pm0.806)\times 10^{-4}$ & $-135 \pm 43.2$ & $(1.96\pm0.532)\times 10^{4}$ &		 $0.83$ & $-1$ & $-0.83$\\
11 & $(4.83\pm2.99)\times 10^{-5}$ & $-66.2 \pm 124$ & $(9.7\pm6.01)\times 10^{4}$ &		 $0.32$ & $-1$ & $-0.32$\\
12 & $(1.71\pm1.57)\times 10^{-5}$ & $-114 \pm 276$ & $(1.88\pm1.73)\times 10^{5}$ &		 $0.37$ & $-1$ & $-0.37$\\
13 & $(1.39\pm31.2)\times 10^{-5}$ & $-163 \pm 3.64\times 10^{3}$ & $(1.21\pm27.3)\times 10^{5}$ &	 $1$ & $-1$ & $-1$\\
14 & $(4.32\pm219)\times 10^{-6}$ & $-1.59 \pm 689$ & $(1.35\pm68.1)\times 10^{5}$ &		 $-0.012$ & $-1$ & $0.012$\\
15 & $(2.66\pm6.97)\times 10^{-5}$ & $-0.0026 \pm 32.2$ & $(3.53\pm9.29)\times 10^{3}$ &		 $-0.091$ & $-1$ & $0.09$\\
16 & $(2.6\pm1.8)\times 10^{-5}$ & $-0.0018 \pm 7.38$ & $144 \pm 109$ &		 $-0.38$ & $-0.98$ & $0.36$\\
\bottomrule[1pt]
	\end{tabular}
	\caption{Same as Tab.~\ref{tab:fit-coeffs-14} but for the FCC with $\sqrt{s}=100$ TeV.}
	\label{tab:fit-coeffs-100}
\end{table}